\documentclass[%
aip,
amsmath,amssymb,
reprint,%
]{revtex4-1}

\usepackage{graphicx}
\usepackage{dcolumn}
\usepackage{bm}

\usepackage[utf8]{inputenc}
\usepackage[T1]{fontenc}
\usepackage{mathptmx}
\usepackage{etoolbox}

\makeatletter
\def\@email#1#2{%
	\endgroup
	\patchcmd{\titleblock@produce}
	{\frontmatter@RRAPformat}
	{\frontmatter@RRAPformat{\produce@RRAP{*#1\href{mailto:#2}{#2}}}\frontmatter@RRAPformat}
	{}{}
}%
\makeatother
\begin{document}
	
	\preprint{AIP/123-QED}
	
	\title[]{Social norms of fairness with reputation-based role assignment in the dictator game}
	\author{Qing Li}

	\author{Songtao Li}%
	
	\author{Yanling Zhang}
	\altaffiliation{yanlzhang@ustb.edu.cn}
	
	\affiliation{
		Key Laboratory of Knowledge Automation for Industrial Processes of Ministry of Education, School of Automation and Electrical Engineering, University of Science and Technology Beijing, Beijing 100083, China
	}%
	
	\author{Xiaojie Chen}
	\altaffiliation{xiaojiechen@uestc.edu.cn}
	\affiliation{%
		School of Mathematical Sciences, University of Electronic Science and Technology of China, Chengdu 611731, China
	}%
	
	\author{Shuo Yang}
	\affiliation{
		Key Laboratory of Knowledge Automation for Industrial Processes of Ministry of Education, School of Automation and Electrical Engineering, University of Science and Technology Beijing, Beijing 100083, China
	}%
	
	\date{\today}
	
\begin{abstract}
A vast body of experiments share the view that social norms are major factors for the emergence of fairness in a population of individuals playing the dictator game (DG). Recently, to explore which social norms are conducive to sustaining cooperation has obtained considerable concern. However, thus far few studies have investigated how social norms influence the evolution of fairness by means of indirect reciprocity. In this study, we propose an indirect reciprocal model of the DG and consider that an individual can be assigned as the dictator due to its good reputation. We investigate the `leading eight' norms and all second-order social norms by a two-timescale theoretical analysis. We show that when role assignment is based on reputation, four of the `leading eight' norms, including stern judging and simple standing, lead to a high level of fairness, which increases with the selection intensity. Our work also reveals that not only the correct treatment of making a fair split with good recipients but also distinguishing unjustified unfair split from justified unfair split matters in elevating the level of fairness.
\end{abstract}
	
	\maketitle
	
\begin{quotation}
Fair behavior among unrelated individuals has been found in a quantity of DG experiments. In the DG, the dictator is assigned to unilaterally divide a given resource between her/himself and the recipient, who unconditionally accepts the division. Making a fair split is remarkable because it is costly for the dictator, but it is beneficial for the recipient. As such, making a fair split is in contradiction with the prediction of standard game theory, which suggests that entirely rational dictators should dispense nothing to recipients. Evolutionary game theory provides a theoretical framework for explaining the mismatch between experimental observations and game theory. The explanation of the mismatch has received ample attention in the recent past. In our paper, we extend the scenario of the random role assignment in the classic evolutionary DG model and take into account the reputation-based role assignment, namely, an individual with good reputation plays the role of dictator when her/his opponent is bad. Our research reveals that the reputation-based role assignment can lead to a high level of fairness for four of the `leading eight' norms, including stern judging and simple standing, and that the level of fairness increases with the selection intensity. These results can provide some insights for a better understanding of the emergence of fairness in the realistic scenario of reputation.
\end{quotation}
	
\section{\label{sec:introduction}Introduction}	
A quantity of canonical experiments using economic games demonstrated that people tend to exhibit prosocial behaviors even in one-shot anonymous interactions~\cite{Camerer2011}. Fair behavior is one of the most important prosocial behaviors in human society, and it is an important undertaking for many social dilemmas, such as accessibility dilemma of antibiotics~\cite{Chen2018} and stalemates in climate talks~\cite{Wang2010}. However, how to understand the emergence of fair behavior among unrelated individuals remains a challenge. As one typical paradigm, the DG has been often used to study the evolution of fair behavior~\cite{Forsythe1994}. In the game, the recipient must accept whatever the dictator delivers, and then the choice of the dictator is not influenced by the retributive motive of the recipient~\cite{Zhang2019}.

A large DG experiment~\cite{Henrich2010}, which was administered across $15$ diverse societies, showed that dictators offered, on average, $37\%$ of the total resource to recipients with a range from $26\%$ to $47\%$. However, a meta-analysis of DG experiments~\cite{Engel2011} found a mean offer of $28\%$. Those results deviate significantly from the prediction by game theory that entirely rational dictators (exclusively driven by the maximization of their own monetary payoffs) ought to dispense nothing to recipients. One classic explanation for the mismatch between the theoretical prediction and experimental observations is about social preference~\cite{Capraro2021}, under which an individual's utility function depends on not only his own monetary payoff, but also the monetary payoff of the other one involved in the DG, e.g., the inequity aversion utility~\cite{Fehr1999}.

However, a series of experimental findings cannot be explained by any utility functions that are based solely on monetary outcomes~\cite{Vostroknutov2020}, and thus the explanation about social preferences is criticized. Despite the fact that the monetary payoff outcome of giving or keeping is identical when playing the DG is compulsory and optional, giving is more likable than keeping in the former and the verse holds in the latter~\cite{Dana2006,Lazear2012}. A similar result was also found when the choice set of dictators is extended~\cite{Bardsley2008}. These results mentioned above imply that beyond the social preference, fair behavior reflects an intrinsic desire to adhere to social norm~\cite{Capraro2019,Fehr2018}, which is defined as `commonly known standards of behavior that are based on widely shared views how group members ought to behave in a given situation'.

Evolutionary game theory, where one strategy with higher payoff is more likely to spread among the population~\cite{Su2022a,Hu2021,Szolnoki2021,Perc2017,Fu2017}, provides a theoretical framework for studying the effect of social norm on the evolution of fairness. Under this framework, a social norm is usually enforced by a reputation system, where the cost of complying with the social norm can be efficiently reduced~\cite{Castelfranchi1998}. Here, the social norm works as a top-down mechanism that impacts the bottom-up behaviors indirectly by generating a reputation uplift/downgrade, underlying indirect reciprocity~\cite{Santos2018socialnorm}. Note that indirect reciprocity has been found to be a fundamental mechanism for the evolution of cooperation in the donation game~\cite{Santos2018,Clark2020,Schmid2021}. Then, a natural question is how fair behavior in the DG is influenced by indirect reciprocity.

In this work, we thereby address the emergence and maintenance of fairness by an indirect reciprocal model. We consider the random role assignment for individuals when they play the DG with each other. Furthermore, note that in the real society, a person with good reputation is more likely to volunteer his time to work at an NGO, donate money to charity, or give money to a homeless person on the street. Thus, besides the random role assignment, we also investigate a way of reputation-based role assignment for individuals. In addition, the widely investigated social norms for the evolution of cooperation use the first-order information (only the action), the second-order information (the recipient's reputation and the action), or the third-order information (the two participants' reputation and the action) to assess the actor's reputation. In an exhaustive research of the third-order social norms~\cite{Ohtsuki2004how}, eight social norms, called as the `leading eight' norms, were found to maintain cooperation. In this paper, we concentrate on the `leading eight' norms together with all possible second-order social norms, and study which social norms can promote the evolution of fairness in our indirect reciprocal model.

\section{\label{sec:model}Model}

\begin{figure}
	\centering
	\includegraphics[width=\linewidth]{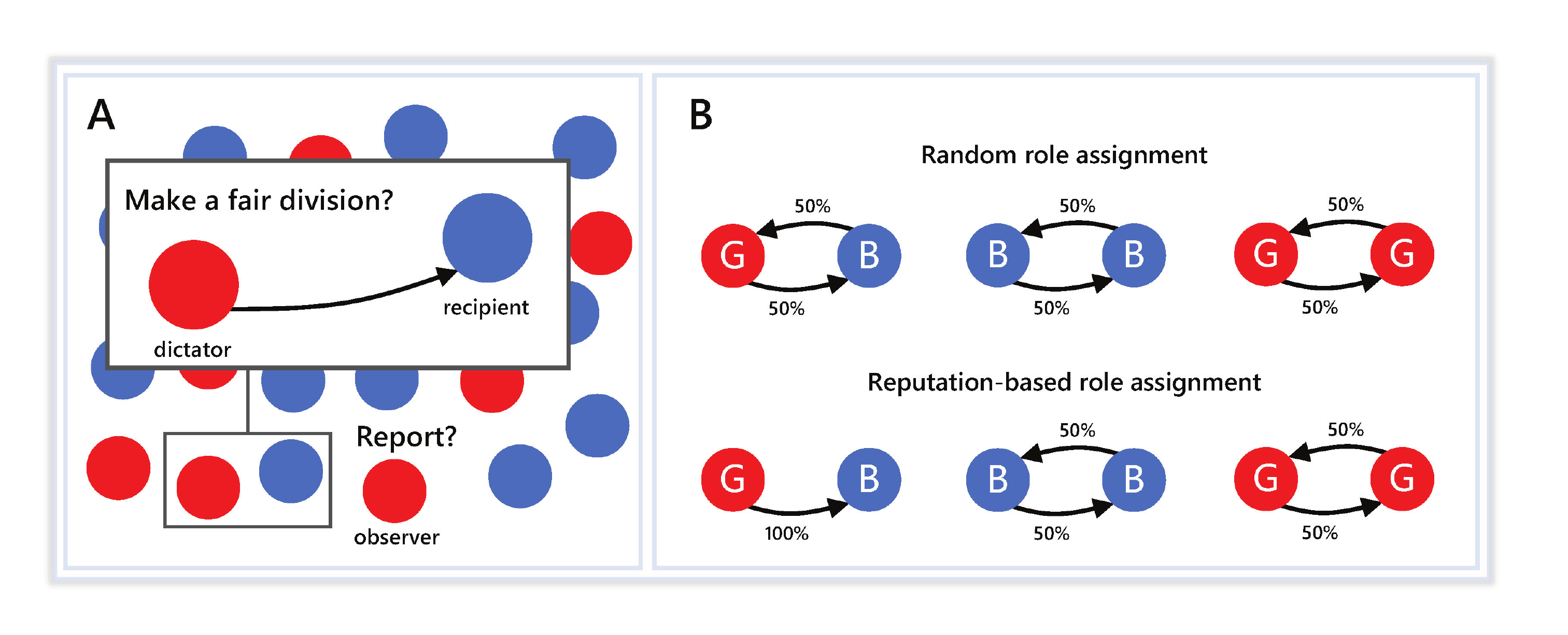}
	\caption{\label{model} The DG under indirect reciprocity. (A) Pairs of players are randomly chosen from the well-mixed population to play the DG, in which the dictator can make a $50-50$ division or a $100-0$ division with a recipient. Then a third player is randomly chosen as the observer to report the dictator's reputation or not. (B) Two ways of role assignment. In the random role assignment, the two participants have the same possibility of becoming the dictator. Yet in the reputation-based role assignment, the good plays the role of dictator when he interacts with the bad; two players with the same reputation are randomly chosen as the dictator or the recipient. }
\end{figure}

We consider a finite well-mixed population with population size $Z$ and each player in the population is assigned with a binary reputation, good or bad. Any two players are randomly chosen from the population to engage in a DG, where one is the dictator and the other one is the recipient. The $50-50$ division is widely observed in economic environments of the real world and the laboratory~\cite{Andreoni2009}.
Accordingly we consider a simplified version of the DG, where the dictator only has two optional strategies, $50-50$ division (fair split is abbreviated as $F$) and $100-0$ division (unfair split is abbreviated as $N$). Since the DG includes two roles of dictator and recipient, two ways of role assignment will be investigated, that is, the random role assignment and the reputation-based role assignment (see the illustrative plot in Fig.~\ref{model}). In the random role assignment, the two participants have the equal possibility of becoming the dictator. While in the reputation-based role assignment, the individual with good reputation plays the role of dictator when the opponent is a bad individual; two players with the same reputation are randomly chosen as the dictator or the recipient. In addition, each interaction can be witnessed by a randomly chosen third player. After observing and assessing the interaction, the observer chooses to report the outcome (abbreviated as $R$) or to be silent (abbreviated as $S$). Note that reporting means that the observer shares the dictator's reputation with all other players, and accordingly bears a personal cost $c_R$.

Here, we denote the strategy of each player in the game by a three-letter string $s_Gs_Bs_R$, which depicts (1) whether the player makes a fair division with a good recipient $(s_G=F)$ or not $(s_G=N)$ if he finds himself in the role of dictator, (2) whether the player makes a fair division with a bad recipient $(s_B=F)$ or not $(s_B=N)$ if in the role of dictator, and (3) whether the player reports $(s_R=R)$ or not $(s_R=S)$ if in the role of observer. Following the common practice~\cite{Suzuki2007}, an implementation error is also considered, meaning that a dictator fails to act fairly when he intends to be fair with probability $\varepsilon$. The observer uses the social norm to assess the dictator's reputation. In this work, we consider the `leading eight' norms and all second-order social norms. Assume that a third-order social norm is represented by two four-dimensional vectors $S^G=(F_G, F_B, N_G, N_B)$ and $S^B=(F_G, F_B, N_G, N_B)$, which denote two cases where the previous reputation of the dictator is good or bad. Regarding each entry of $S^G$ and $S^B$, we set $L_M=1$, which means that the dictator is labelled with a good credit when he takes action $L$ against a recipient of reputation $M$, and analogously $L_M=0$, which means that the corresponding dictator is regarded as bad.
Particularly, a second-order social norm satisfies $S^G=S^B$.

After individuals play the game and obtain the payoffs, they will take the strategy imitation by the pairwise comparison rule~\cite{Traulsen2007}. To be specific, in each generation two players are randomly chosen to become the focal one and the model one, respectively. With a small probability $\mu$, a mutation occurs, meaning that the focal one randomly adopts one of all candidate strategies. Otherwise (with probability $1-\mu$), the focal one $x$ has an opportunity of imitation, and he imitates the strategy of the model one $y$ with the following probability $$p(x\to y) = 1/(1+e^{-\beta(g_y-g_x)}),$$
where $\beta$ denotes the intensity of selection, $g_x$ and $g_y$ are the payoffs of $x$ and $y$, respectively. The $\beta$ stands for how closely the strategy dynamics relies on the outcome of interactions~\cite{Du2018,Su2022b,Szolnoki2014}. For weak selection $\beta \to 0$, the update slightly depends on the payoffs of individuals and randomness matters in the strategy selection.

\section{\label{sec:methods}Methods}
In this study, we assume that the time scale for strategy selection is much slower than the one for reputation update~\cite{Santos2018socialnorm,87Hilbe2018}. Accordingly, we first calculate the steady-state distribution of the reputation system and the expected payoff of each strategy by considering that the strategies of all players are fixed. We then compute the fixation probability between any pair of strategies and the steady-state frequency of each strategy by allowing players to update their strategies over time.

Given that the strategies of all players are fixed, the reputation system can be described by a Markov chain. When a population consists of $m$ players with strategy $X=s_G^Xs_B^Xs_R^X$ and $Z-m$ players using strategy $Y=s_G^Ys_B^Ys_R^Y$, the state of the reputation system is denoted by a two-dimensional vector $(i,j)$, which means that there are $i\in {0,1,\cdots, m}$ ($j\in {0,1,\cdots, Z-m}$) good players among $m$ $X$-players (among $Z-m$ $Y$-players). Given that the reputation state is $(i,j)$ at time $t$, we can calculate the transition probability $P(i,j;i',j')$, which characterizes how likely the population will be in state $(i',j')$ at time $t+1$ (see Appendix~\ref{reputation} for the detailed calculation).

Let $V=(v(i,j))$ be the steady-state distribution of the reputation system with the transition matrix $P=(P(i,j;i',j'))$, and accordingly we have $VP=V$ and $\begin{matrix} \sum_{i,j} v(i,j)=1 \end{matrix}$. Here each entry $v(i,j)$ means the expected frequency of the state $(i,j)$ which we observe over the course of the reputation dynamics. Note that the expected payoffs of $X$ and $Y$ in a population with $m$ $X-$players and $Z-m$ $Y-$players, $g_X(m)$ and $g_Y(m)$, are respectively calculated as
\begin{eqnarray}\label{expected payoffs} 
\begin{array}{l}
g_X(m) =\sum_{i=0}^m{\sum_{j=0}^{Z-m}{v( i,j ) \pi_X( m,i,j )}},\\
g_Y(m) =\sum_{i=0}^m{\sum_{j=0}^{Z-m}{v( i,j ) \pi_Y( m,i,j )}},
\end{array}
\end{eqnarray}
where $\pi_X( m,i,j )$ and $\pi_Y( m,i,j )$ are the expected payoffs of $X$ and $Y$ when there are $i$ good players among $m$ $X$-players and $j$ good players among $Z-m$ $Y$-players (see Appendix~\ref{payoff} for their expressions).

We now focus on how players change their strategies over time, which occurs on a much slower time scale than the reputation update. In the limit of low mutation $\mu\to0$, the strategy dynamics can be described by an embedded Markov Chain~\cite{Fudenberg2006}. The state space includes eight monomorphic populations, each of which adopts one of $FFR$, $FFS$, $FNR$, $FNS$, $NFR$, $NFS$, $NNR$, and $NNS$. The corresponding transition matrix is $A=(a_{XY})_{8\times8}$, where $a_{XY}$ can be expressed by the fixation probability between any pair of strategies (see Appendix~\ref{fixation} for the detailed expression). The normalised left eigenvector $\Phi$ of the stochastic matrix $A$ associated with the eigenvalue $1$ satisfies $\Phi=\Phi A$. Here $\Phi=(\phi_{FFR},\phi_{FFS},\phi_{FNR},\phi_{FNS},\phi_{NFR},\phi_{NFS},\phi_{NNR},\phi_{NNS})$ is the steady-state distribution of the strategy dynamics, where $\phi_X$ is the fraction of time spent in each monomorphic population of $X$. Accordingly by assuming that $f_F(X)$ is the fairness level of the monomorphic population of $X$, we define the total level of fairness $F_F$ as the weighted average of $f_F(X)$ with the weight $\phi_X$,
\begin{eqnarray}\label{f_F}
F_F=\sum_{X} \phi_Xf_F(X).
\end{eqnarray}

In the monomorphic population of $X$, we assume that $p_F(i|X)$ is the probability for a player to make a fair division in the role of dictator or to receive a fair division in the role of recipient when there are $i$ good players. Then $f_F(X)$ is the weighted average of $p_F(i|X)$ with the weight $v(i|X)$, which is the proportion of time spent in a state of $i$ good players in a monomorphic population of $X$. Accordingly, $f_F(X)$ is given as
\begin{eqnarray}
f_F(X)=\sum_{i} v(i|X)p_F(i|X).
\end{eqnarray}
For unconditionally fair/unfair strategies, $f_F(X)$ is independent of $v(i|X)$, i.e., $f_F(X)=p_F(i|X)$ for $X=NNS$, $NNR$, $FFS$, or $FFR$. Then, we have $f_F(X)=0$ for $X=NNS$ or $NNR$ and $f_F(X)=1-\varepsilon$ for $X=FFS$ and $FFR$ because an $X$ dictator is always unfair or always fair except the implementing error, irrespective of the opponent's reputation. In a monomorphic population of $NFS$ or $FNS$, individuals cannot obtain the reputation information of opponents. Then we assume that they act fairly with a fixed probability and let the probability be only $0.1$ for simplicity. Therefore, we have $f_F(X)=p_F(i|X)=0.1$ for $X=NFS$ and $FNS$, suggesting that $NFS$ and $FNS$ have little effect on the total level of fairness. Different from the above six strategies, $f_F(FNR)$ and $f_F(NFR)$ depends closely on $v(i|X)$, which is equal to $v(i,0)$ in a monomorphic population of $X$.
In addition, according to the definition of $p_F(i|X)$, we have $p_F(i|X)$ for $X=NFR$ or $FNR$ as follows
\begin{eqnarray}
\begin{array}{lll}
p_F(i|X)&=&(1-\epsilon)I(s_G^X)\frac{i}{Z}+(1-\epsilon)I(s_B^X)\frac{Z-i}{Z},\\ &&\mbox{when role assignment is random,}\\
p_F\left( i|X \right) &=&\left( 1\!-\!\varepsilon \right) \left( \frac{i\left( i-1 \right)}{Z\left(Z-1 \right)}I\left( s_{G}^{X} \right)\! +\!\frac{Z^2-Z-i^2+i}{Z\left(Z-1 \right)}I\left( s_{B}^{X} \right) \right),\\ &&\mbox{when role assignment is based on reputation,}
\end{array}
\end{eqnarray}
where $I(x)=1$ for $x=F$ and $I(x)=0$ for $x=N$.
	
\section{\label{sec:results}Results}
	
\subsection{Evolutionary outcomes for four typical social norms}

\begin{figure}
	\centering
	\includegraphics[width=\linewidth]{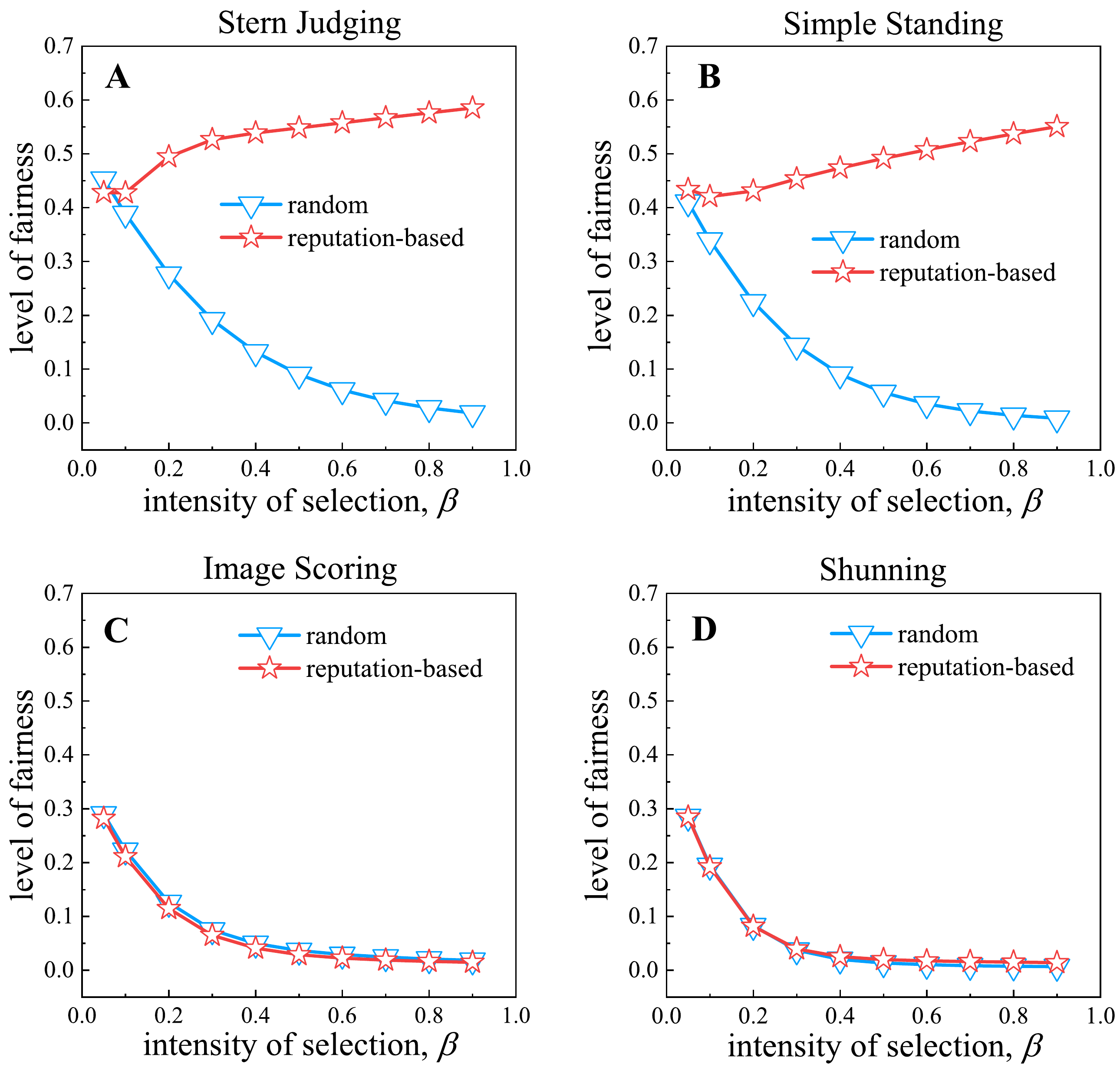}
	\caption{\label{fairness level} The level of fairness as a function of the selection intensity $\beta$ under stern judging (A), simple standing (B), image scoring (C), and shunning (D).  Parameters: $Z=50$, $\varepsilon=0.01$, $c_R=0.01$, and $\mu=0.01$.}
\end{figure}

We study the level of fairness by numerically calculating the stationary distribution of the strategies and the one of the reputation system in each monomorphic population. We first focus on four social norms most frequently studied in previous work about indirect reciprocity~\cite{Radzvilavicius2021}: stern judging $S^G=S^B=(1,0,0,1)$, simple standing $S^G=S^B=(1,1,0,1)$, image scoring $S^G=S^B=(1,1,0,0)$, and shunning $S^G=S^B=(1,0,0,0)$. We consider both the random role assignment and the reputation-based role assignment for each social norm. Fig.~\ref{fairness level} shows the level of fairness as a function of the selection intensity $\beta$. Under stern judging and simple standing (Figs.~\ref{fairness level}A and~\ref{fairness level}B), the level of fairness for the reputation-based role assignment is similar to the one for the random role assignment at small $\beta$, and becomes significantly higher than the one for the random role assignment at large $\beta$. Here, the level of fairness increases with $\beta$ when role assignation is based on reputation, and it decreases with $\beta$ when role assignment is random. Whereas under image scoring and shunning (Figs.~\ref{fairness level}C and~\ref{fairness level}D), the level of fairness for the reputation-based role assignment is almost identical with the one for the random role assignment irrespective of $\beta$. Here, the level of fairness falls sharply to around zero as $\beta$ increases in the two ways of role assignment.

\begin{figure}
	\centering
	\includegraphics[width=\linewidth]{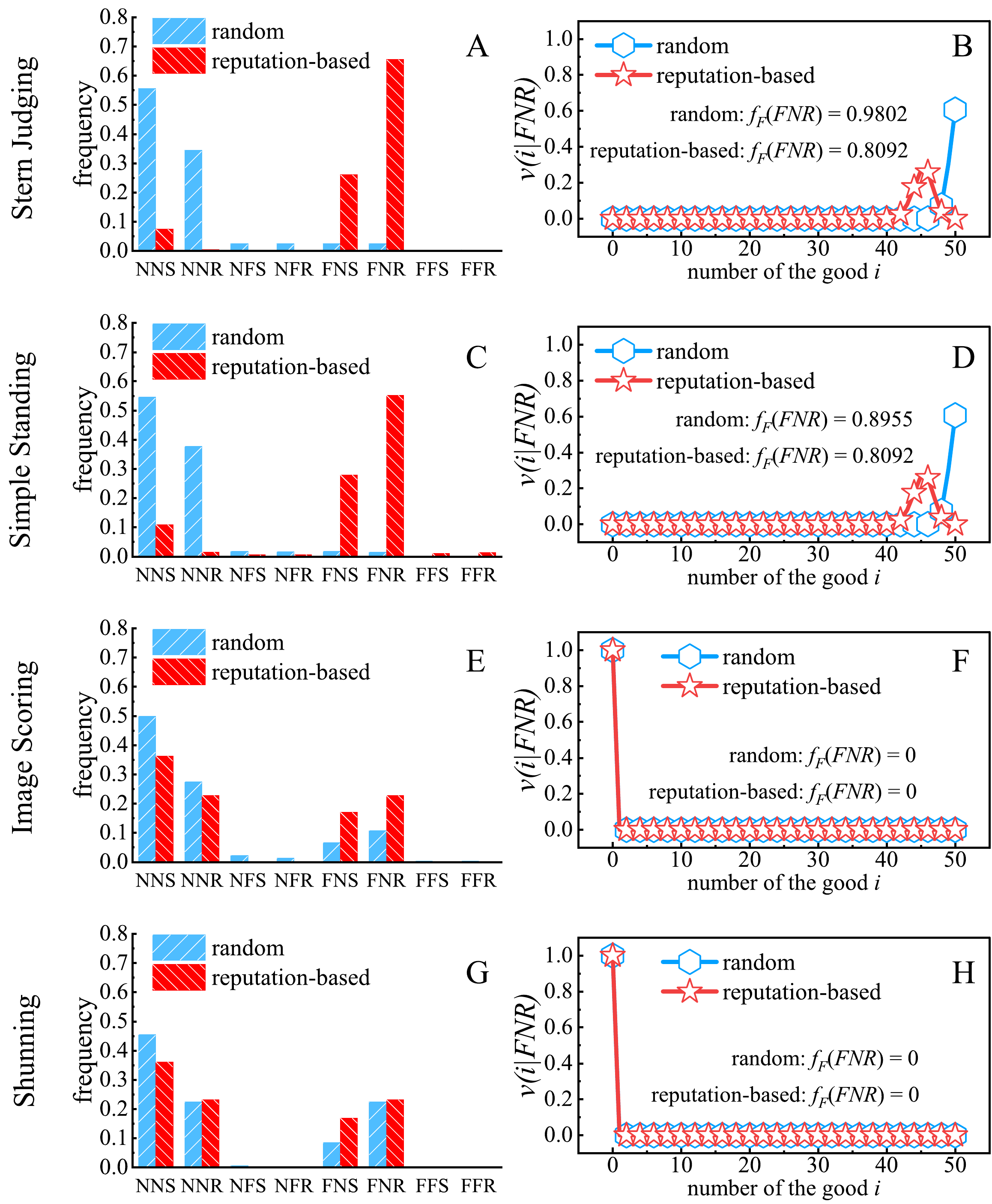}
	\caption{\label{strategy} Under four representative social norms, the time averaged frequency of each strategy during the evolutionary process (A, C, E, G) and the reputation dynamics when all players adopt $FNR$ (B, D, F, H). Parameters: $Z=50$, $\varepsilon=0.01$, $c_R=0.01$, $\mu=0.01$, and $\beta=0.6$.}
\end{figure}

Indeed the phenomenon mentioned above can be understood from the perspective of the steady-state frequency $\Phi(X)$ and the fairness level $f_F(X)$ of each monomorphic population (see Eq.~(\ref{f_F})). We show the steady-state frequency of each strategy in Fig.~\ref{strategy}. Two unconditionally fair strategies, $FFR$ and $FFS$, have diminishing steady-state frequencies at large $\beta$ ($\beta=0.6$), because they cannot resist the invasion of $NNR$ and $NNS$. Then they contribute little to the level of fairness although the monomorphic population of $FFR$ or $FFS$ shows a high fairness level ($f_F(X)=1-\varepsilon$). Similar to $FFR$ and $FFS$, $NFR$ and $NFS$ also contribute little to the level of fairness because of their low frequencies. The steady-state frequencies of $FNR$ and $FNS$ differ significantly with social norm or role assignment, suggesting they can obviously enlarge the level of fairness.
Therefore, whether fairness can evolve depends closely on $FNR$ and $FNS$.

In particular, under stern judging and simple standing, the monomorphic population of $FNS$ exhibits a low fairness level
($f_F(X)=0.1$), and the frequency of $FNS$ is at most $0.3$.
Moreover, the monomorphic population of $FNR$ exhibits similar high fairness levels ($>80\%$) in the two ways of role assignment (Figs.~\ref{strategy}B and~\ref{strategy}D). Accordingly, $FNR$ is the key strategy for explaining whether fairness emerges. Here, $FNR$ has a much higher frequency for the reputation-based role assignment than the random role assignment (Figs.~\ref{strategy}A and~\ref{strategy}C), suggesting that the reputation-based role assignment leads to a higher level of fairness than the random role assignment. As $\beta$ increases, $FNR$ becomes extinct in the population when role assignment is random, implying that the level of fairness decreases with $\beta$, but $FNR$ gradually predominates in the population when role assignment is based on reputation, meaning
that the level of fairness increases with $\beta$.

We can understand the reputation system for the monomorphic population of $FNR$ as follows.
If the implementation error is absent, the state of `all good' will be the absorbing state for stern judging and simple standing irrespective of role assignment. However, when a nonzero implementation error is involved, the corresponding reputation dynamics no longer has the absorbing states, because bad players can appear in the state of `all good' by mistakenly acting unfairly against a good opponent. As shown in Figs.~\ref{strategy}B and~\ref{strategy}D, the stationary distributions peak at the state of `all good' when role assignment is random, but peak at a mixed state when role assignment is reputation-dependent. In the former, a bad $FNR$ player can obtain good reputation with probability $0.5(1-\epsilon)$ as the dictator when his opponent is a good $FNR$ player. Yet in the latter, a bad $FNR$ player no longer has opportunities of obtaining good reputation when his opponent is a good $FNR$ player (the bad one is the recipient and the good one is the dictator).
For image scoring and shunning, the reputation system has an absorbing state of `all bad' (Figs.~\ref{strategy}F and~\ref{strategy}H) because a $FNR$ player cannot be assessed as good in the state `all bad' by always acting unfairly against bad opponents.

\begin{figure}
	\centering
	\includegraphics[width=\linewidth]{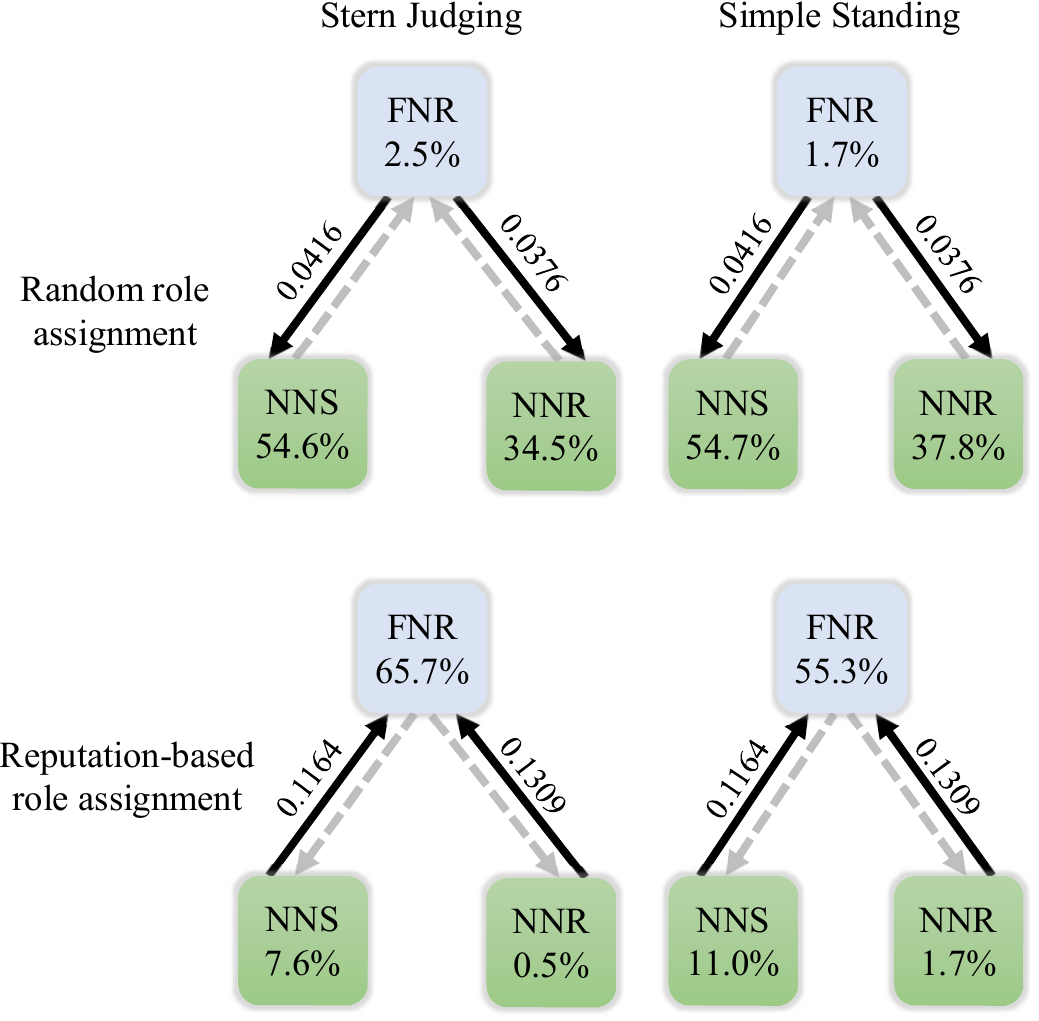}
	\caption{\label{fix} The pairwise competition between $FNR$ and $NNS$/$NNR$ for stern judging and simple standing. The letters and the numbers denote the strategies and their frequencies in the selection-mutation equilibrium. The numbers close to the arrows denote the fixation probability of a single mutant (the ending point) into the given resident strategy (the starting point). The solid lines for the arrows are used to show the fixation probability which is more than the neutral probability $1/Z$. The dashed lines for the arrows means that the fixation probability is less than $1/Z$. Parameters: $Z=50$, $\varepsilon=0.01$, $c_R=0.01$, and $\beta=0.6$.}
\end{figure}

To further understand the frequency of $FNR$, we show the pairwise competition between $FNR$ and $NNS/NNR$, presented in Fig.~\ref{fix}.
Here, we first address some necessary definitions.
If the fixation probability of a single $X$ mutant into the given resident strategy $Y$ is more (less) than the neutral probability $1/Z$, then $X$ can (not) invade $Y$. For the random role assignment, $FNR$ cannot invade $NNS$ or $NNR$, but $NNS$ or $NNR$ can invade $FNR$, causing $FNR$ to have very low frequencies.
Yet for the reputation-based assignment, the $FNR$ players, who deserve good reputation under stern judging and simple standing, earn more chances of being dictators. Therefore, $FNR$ can invade $NNS$ or $NNR$, and yet $NNS$ or $NNR$ cannot invade $FNR$, leading $FNR$ to be played in substantial time.
Moreover the competitive advantage of $FNR$ over $NNS$ or $NNR$ is enlarged with growing $\beta$, meaning that the larger $\beta$ leads to a higher level of fairness.

On the other hand, under image scoring and shunning,
$FNR$ and $FNS$ have slightly higher frequencies for the reputation-based role assignment than the random role assignment (Figs.~\ref{strategy}E and~\ref{strategy}G).
However, we observe that the monomorphic population of $FNR$ has a diminishing probability of acting fairly from Figs.~\ref{strategy}F and~\ref{strategy}H, because the corresponding reputation dynamics has the only absorbing state `all bad'.
Moreover, the monomorphic population of $FNS$ exhibits a low fairness level ($f_F(X)=0.1$).
Accordingly, the reputation-based role assignment leads to almost the same level of fairness as the random role assignment.

\subsection{Evolutionary outcomes for other social norms}
\begin{figure*}
	\centering
	\includegraphics[width=\linewidth]{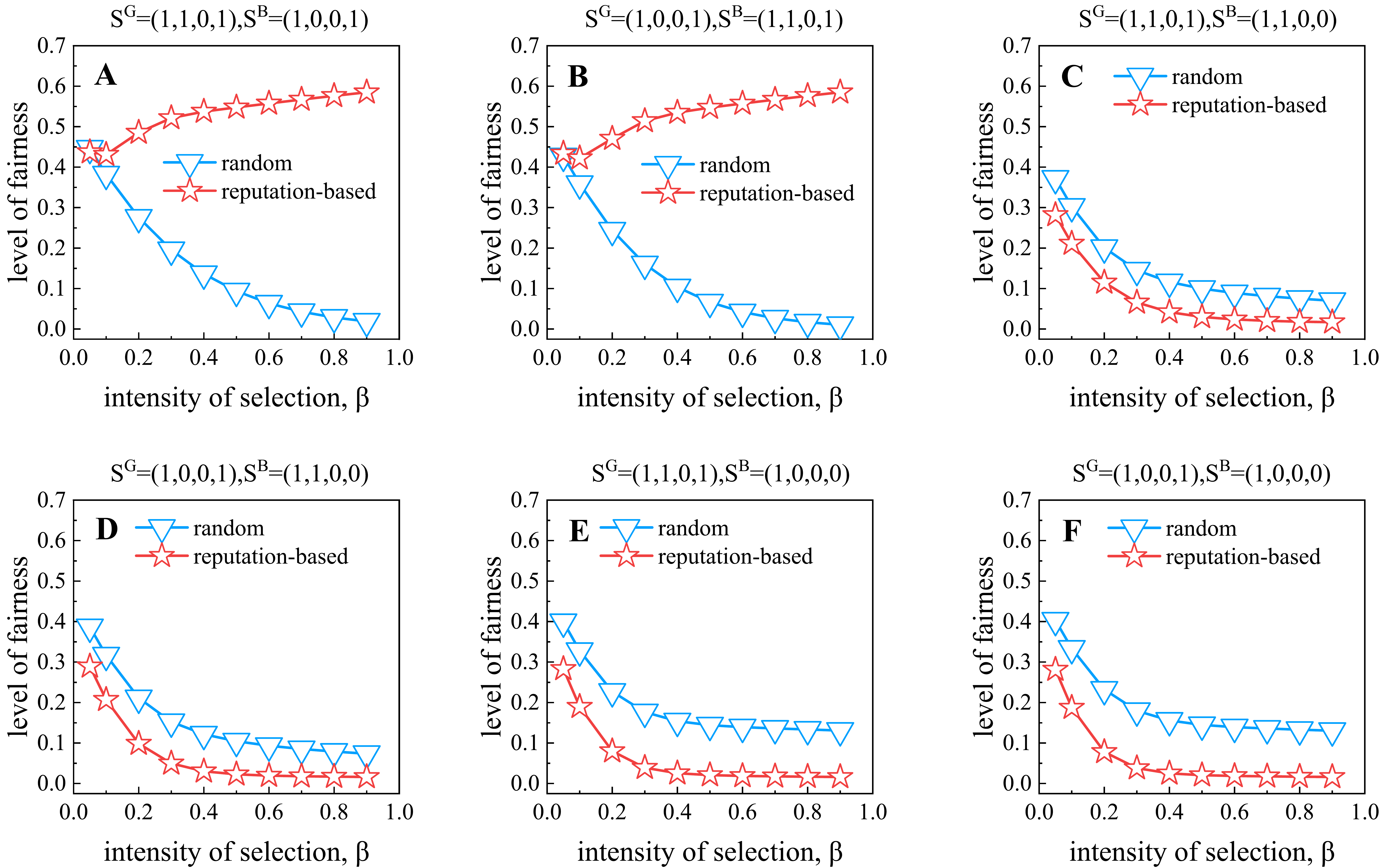}
	\caption{The level of fairness as a function of the intensity of selection under the remaining `leading eight' social norms. For (A) and (B), the $S^G$ and $S^B$ of the social norms are $(1,0,0,1)$ or $(1,1,0,1)$. For (C)-(F), the $S^G$ of the social norms is  $(1,0,0,1)$ or $(1,1,0,1)$, and the $S^B$ is $(1,1,0,0)$ or $(1,0,0,0)$.
	Parameters: $Z=50$, $\varepsilon=0.01$, $c_R=0.01$, and $\mu=0.01$.}
	\label{third}
\end{figure*}

We also consider the remaining `leading eight' norms except stern judging and simple standing. When a social norm with $S^G=(1,1,0,1)$ and $S^B=(1,0,0,1)$ or a social norm with $S^G=(1,0,0,1)$ and $S^B=(1,1,0,1)$ is used (Figs.~\ref{third}A and~\ref{third}B), fairness is allowed to emerge and to increase with $\beta$ for the reputation-based role assignment. Note that $S^G$ and $S^B$ of these two social norms are the corresponding one of stern judging or simple standing.
Then similar to stern judging or simple standing, we have the following conclusions.
When role assignment is based on reputation, $FNR$ is played in substantial time and the monomorphic population of $FNR$ exhibits a high fairness level (Figs.~\ref{thirdrepu}A-\ref{thirdrepu}D), suggesting that fairness can evolve.
However when role assignment is random, $FNR$ is played in little time, implying that fairness cannot evolve.
The frequency of $FNR$ can be understood from the competition between $FNR$ and $NNR/NNS$, which is presented in Fig.~\ref{thirdfix}. When role assignment is based on reputation, $FNR$ is both able to invade $NNR/NNS$ and able to resist the invasion of $NNR/NNS$, inducing $FNR$ to be played in much time.
Yet when role assignment is random, $FNR$ is neither able to invade $NNR/NNS$ nor able to resist the invasion of $NNR/NNS$, leading $FNR$ to be played in little time.

Accordingly, four of the `leading eight' norms, including stern judging and simple standing, can efficiently promote the evolution of fairness for the reputation-based role assignment. The common point of the four social norms lies in the assessment that allocating half (fair split) to a good recipient is regarded as good. Besides this point, all of them distinguish unjustified unfair split from justified unfair split. To be specific, unjustified unfair split downgrades the dictator's reputation, i.e., the dictator who acts unfairly against a good recipient is perceived as bad; justified unfair split elevates the dictator's reputation, i.e., making an unfair split with a bad recipient is regarded as good. This suggests that not only the correct treatment of allocating half (fair split) to a good recipient but also distinguishing unjustified unfair split from justified unfair split matters in favoring the evolution of fairness. Note that for the four social norms, the positive effect on fairness is sensitive to the cost of reporting $c_R$ (Fig.~\ref{cost}). When $\beta$ is large and $c_R$ is absent, the level of fairness is high, but it gradually decreases with $c_R$.

As shown in Figs.~\ref{third}C-~\ref{third}F, the level of fairness under four of the `leading eight' norms decreases with $\beta$ for the random and reputation-based role assignment. It is obvious that $S^G$ of the four social norms is identical with the corresponding one of stern judging or simple standing, and $S^B$ of them is the same as the corresponding one of image scoring or shunning.
Surprisingly, the level of fairness for the random role assignment is a little higher than the one for the reputation-based role assignment. The phenomenon can be understood from Figs.~\ref{thirdrepu}E-\ref{thirdrepu}L.
$FNR$ and $FNS$ ($F_X(FNS)=0.1$) have similarly low frequencies in the two ways of role assignment.
When role assignment is random, the monomorphic population of $FNR$ exhibits a high fairness level more than $60\%$. Yet when role assignment is based on reputation, the monomorphic population of $FNR$ has a diminishing fairness level, whose reputation system has the only single absorbing state `all bad'.
Then, the random role assignment leads to a slightly higher lever of fairness than the reputation-based role assignment.

We also investigate other second-order social norms~(see Fig.~\ref{figA1}). Just like image scoring and shunning, we can find that a certain level of fairness can emerge under weak selection, but quickly diminishes with growing $\beta$ for the random and the reputation-based role assignment.

\section{\label{sec:discussion}Discussions and Conclusions}
Under the framework of evolutionary game theory~\cite{Liu2018,Szolnoki2019,Perc2013}, it is still difficult to explain why individuals offer fair divisions of a resource, although a little headway has been made by the ultimatum game (UG)~\cite{Deng2021,Zheng2022,Zhang2019,Zhang2018}. Indeed the UG is similar to the DG, except that the recipient explicitly has leverage over the proposer through the ability of rejecting offers. If the recipient accepts the offer, the resource is divided as proposed, otherwise both players receive nothing. Compared with the DG, the UG allows the recipient to have a aspiration by rejecting low offers. We have also conducted a similar numerical computation on the UG when role assignment is random  (unshown). We find the level of fairness for the UG is obviously higher than the one of the DG, implying the significantly positive effect of the recipient's aspiration on fairness. In fact, the positive effect of the aspiration has been widely observed in the realm of cooperation~\cite{Wu2018,Chen2008}.

For the UG, fairness could evolve in the well-mixed population~\cite{Rand2013}, because mutation and drift lead to the gradual increase of responders' minimum acceptable offers. However, such a mechanism cannot work in the DG because the recipient cannot reject offers. Thus in the framework of the DG, it is more difficult to explore the evolutionary origin of fairness. To date, limited studies about the DG focus on three types of mechanisms: degree-based role assignation~\cite{Deng2014}, multilevel selection~\cite{Jeffrey2015}, and co-evolutionary dynamics~\cite{Snellman2018}. These studies are performed in structured populations, allowing network reciprocity to be one of the key factors for fairness. To avoid the effects of network reciprocity, in this work we investigate the game in a well-mixed population, and accordingly we can easily explore the underlying mechanism for the evolution of fairness.

A range of recent theoretical studies have indeed demonstrated the importance of role assignment as a mechanism to enforce fair behavior in human society. In most relevant studies, role assignment is based on degree of various heterogeneous networks, and the impact of role assignment on fairness has been performed in the UG~\cite{Li2013,Wu2013} (two-person or multi-player), the DG~\cite{Deng2014}, and the mixture of these two games~\cite{Chen2019}. To explain the evolution of fairness in the UG, role assignment is also combined with partner choice and reputation~\cite{Yang2015,Deng2021}. In this paper, we focus on the effect of the reputation-based role assignment on the DG.

A vast body of behavioral experiments on the DG share the view that social norms are major factors that drive fair behaviors in multi-agent systems~\cite{Capraro2019}. Which social norms are conducive to sustaining cooperation in the donation game has obtained considerable concern in the research of indirect reciprocity~\cite{Fu2008, OKada2020}. However, thus far few studies have investigated how social norms impact fair behaviors in the DG by means of indirect reciprocity. We here develop an indirect reciprocal model of the DG to determine the social norms governing reputation systems, under which natural selection shapes fair behaviors. Irrespective of social norm and role assignment (random or reputation-based), we show that a certain level of fairness can emerge under weak selection. Moreover, the level of fairness induced by certain weak selection intensities matches the observed behavior in lots of experiments~\cite{Engel2011}, mainly due to randomness. The similar result has also been found in the framework of UG~\cite{Rand2013}.

For the random role assignment, as randomness decreases with growing intensity of selection, we show that the level of fairness quickly diminishes, suggesting that reputation alone cannot efficiently maintain fairness. It nicely fits experimental findings~\cite{Beersma2011,Piazza2008}, in which reputation alone seems to be effective only when consequences of reputation are evident. Indeed the role of reputation in the evolution of fairness is initially studied in Ref.~\cite{Nowak2000}, in which responders are endowed with `reputation' according to what responders have accepted and rejected in the past. This study provides a theoretical evidence that `reputation' in general can lead to fairness in the UG. However, a later study concluded that the key factor for the evolution of fairness is that the strategy space is restricted, suggesting that reputation alone cannot promote fairness when the strategy space is not restricted~\cite{Andre2011}.

In this study, to overcome the incapability of reputation, we incorporate the reputation-based role assignment in the DG. We focus on two main questions: 1) Will fairness emerge for strong selection? 2) Which social norms can outcompete in promoting the emergence and maintenance of fairness for strong selection? We find that four of the `leading eight' norms, including stern judging and simple standing, leading a high level of fairness to emerge and to increase with the selection intensity. It suggests that not only the correct treatment of making a fair split with good recipients but also distinguishing unjustified unfair split from justified unfair split plays an important role in favoring fairness.

Our results are obtained by a two-timescale theoretical analysis. We also investigate the scenario where reputation and strategies evolve at the same scale by agent-based simulations. As shown in Fig.~\ref{simu}, we find that the results are qualitatively similar to the one of two separate time scales for the evolution of reputation and strategies. Under stern judging and and simple standing, the level of fairness for the reputation-based role assignment is obviously higher than the one for the random role assignment. The level of fairness is robust to $\beta$ when role assignment is reputation-dependent and is sensitive to $\beta$ when role assignment is random. Under image scoring and shunning, the level of fairness is sensitive to $\beta$ and quickly diminishes with growing $\beta$ irrespective of role assignment.

There is a promising perspective for the future study. It is known that players tend to treat in-group members preferentially~\cite{Masuda2015}. Therefore, our study can be extended to the scenarios involving groups. For example, a player always acts fairly with an in-group member irrespective of reputation; yet he acts fairly with a good out-group member and acts unfairly with a bad out-group member. In the case, how indirect reciprocity involving groups affects the evolution of fairness can be investigated.

\begin{acknowledgments}
This research was supported by the National Natural Science Foundation of China (Grants Nos. 62033010, 61603036, 61976048).	
\end{acknowledgments}

 \appendix

 \begin{figure}
 	\centering
 	\includegraphics[width=\linewidth]{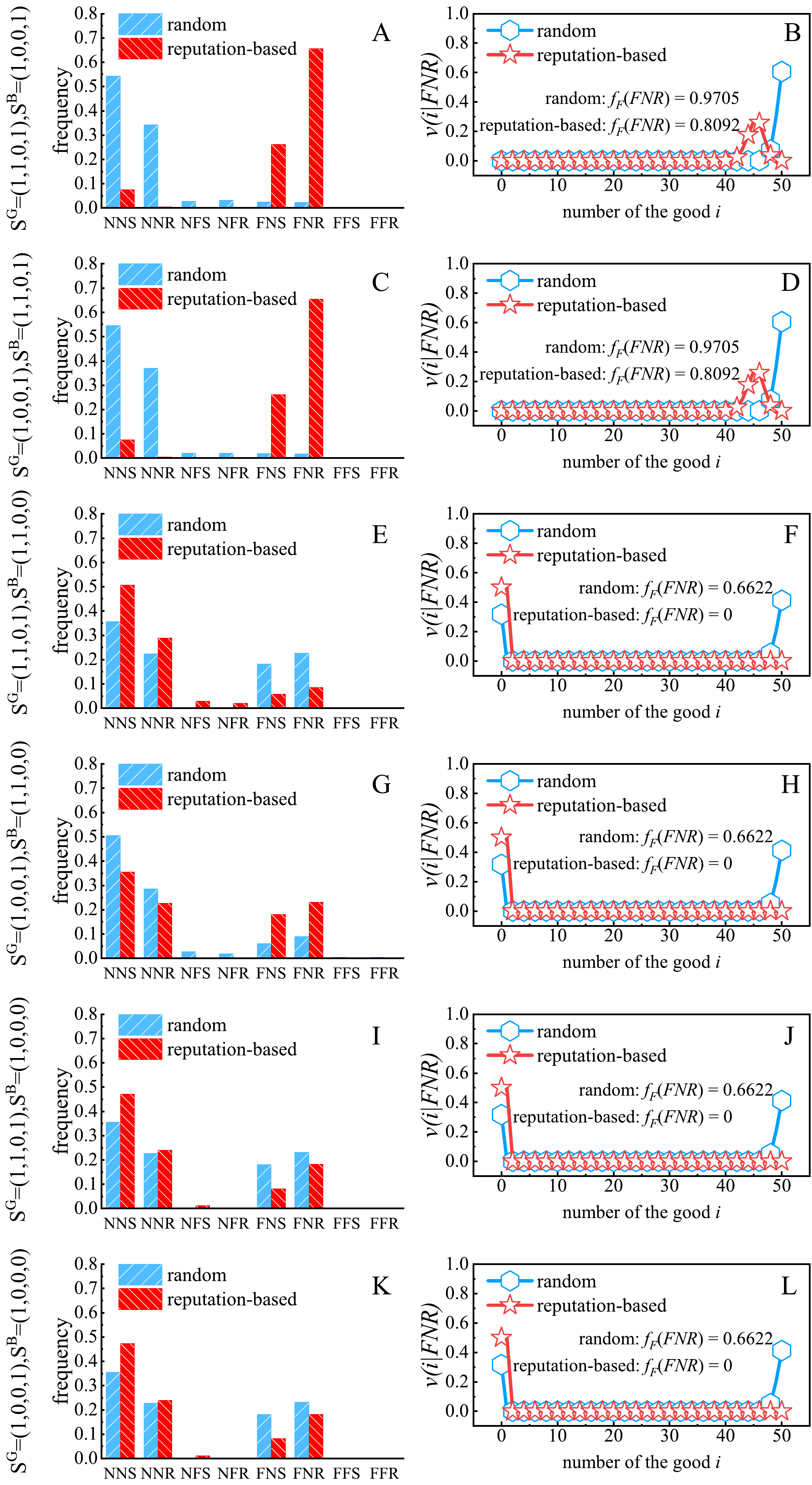}
 	\caption{\label{thirdrepu} Under the remaining `leading eight' social norms, the time averaged frequency of each strategy during the evolutionary process (A, C, E, G, I, K) and the reputation dynamics when all players adopt $FNR$ (B, D, F, H, J, L) . Parameters: $Z=50$, $\varepsilon=0.01$, $c_R=0.01$, $\mu=0.01$, and $\beta=0.6$.}
 \end{figure}

 \begin{figure}
 	\centering
 	\includegraphics[width=\linewidth]{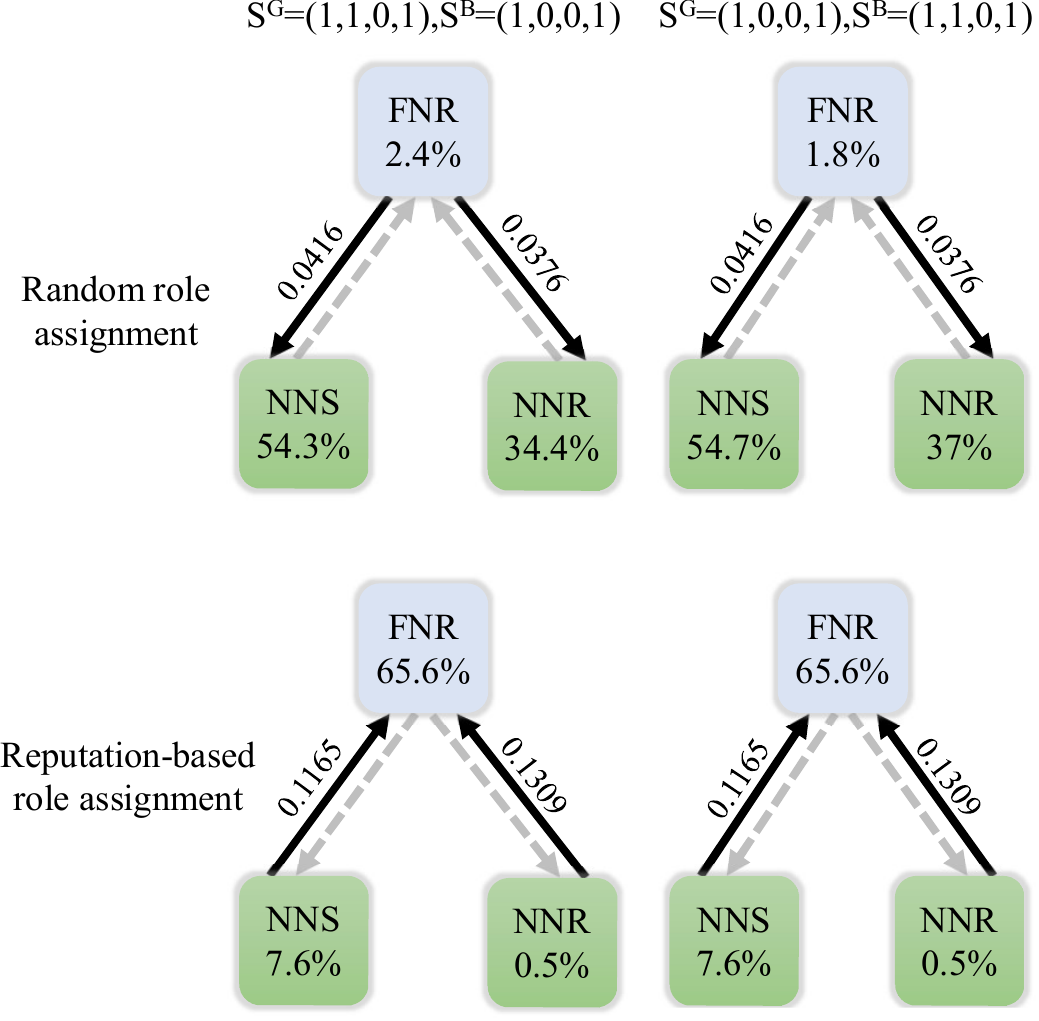}
 	\caption{The pairwise competition between $FNR$ and $NNS$/$NNR$
 		for two social norms whose $S^G$ and $S^B$ $(S^G\neq S^B)$ are $(1,1,0,1)$ or $(1,0,0,1)$. The letters and the numbers denote the strategies and their frequencies in the selection-mutation equilibrium. The numbers close to the arrows denote the fixation probability of a single mutant (the ending point) into the given resident strategy (the starting point). The solid lines for the arrows are used to show the fixation probability which is more than the neutral probability $1/Z$. The dashed lines for the arrows means that the fixation probability is less than $1/Z$. Parameters: $Z=50$, $\varepsilon=0.01$, $c_R=0.01$, and $\beta=0.6$.}
 	\label{thirdfix}
 \end{figure}

\begin{figure}
	\centering
	\includegraphics[width=\linewidth]{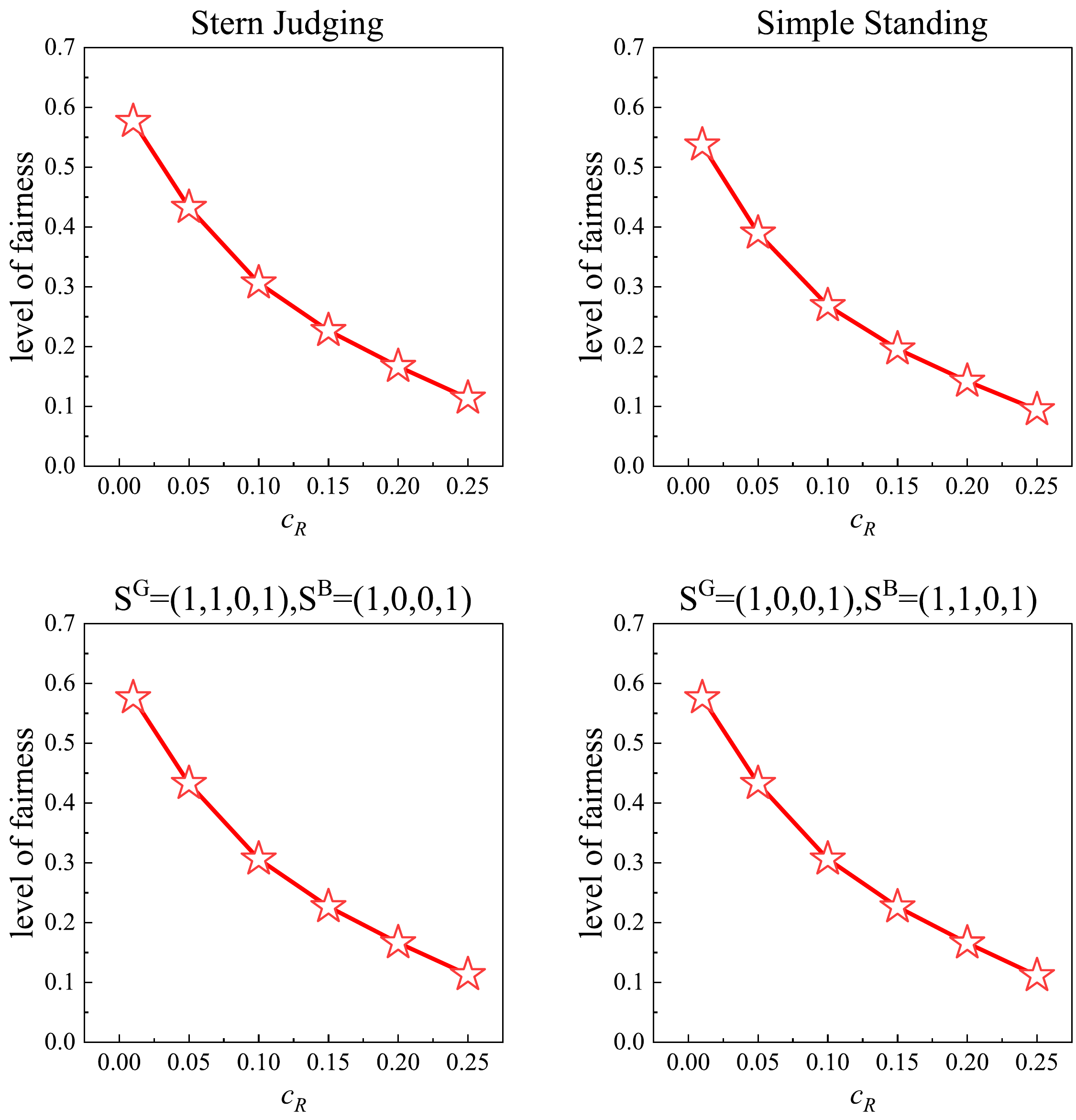}
	\caption{The level of fairness as a function of the cost of resporting. Parameters: $Z=50$, $\varepsilon=0.01$, $\beta=0.8$, and $\mu=0.01$.}
	\label{cost}
\end{figure}

 \begin{figure*}
 	\centering
 	\includegraphics[width=\linewidth]{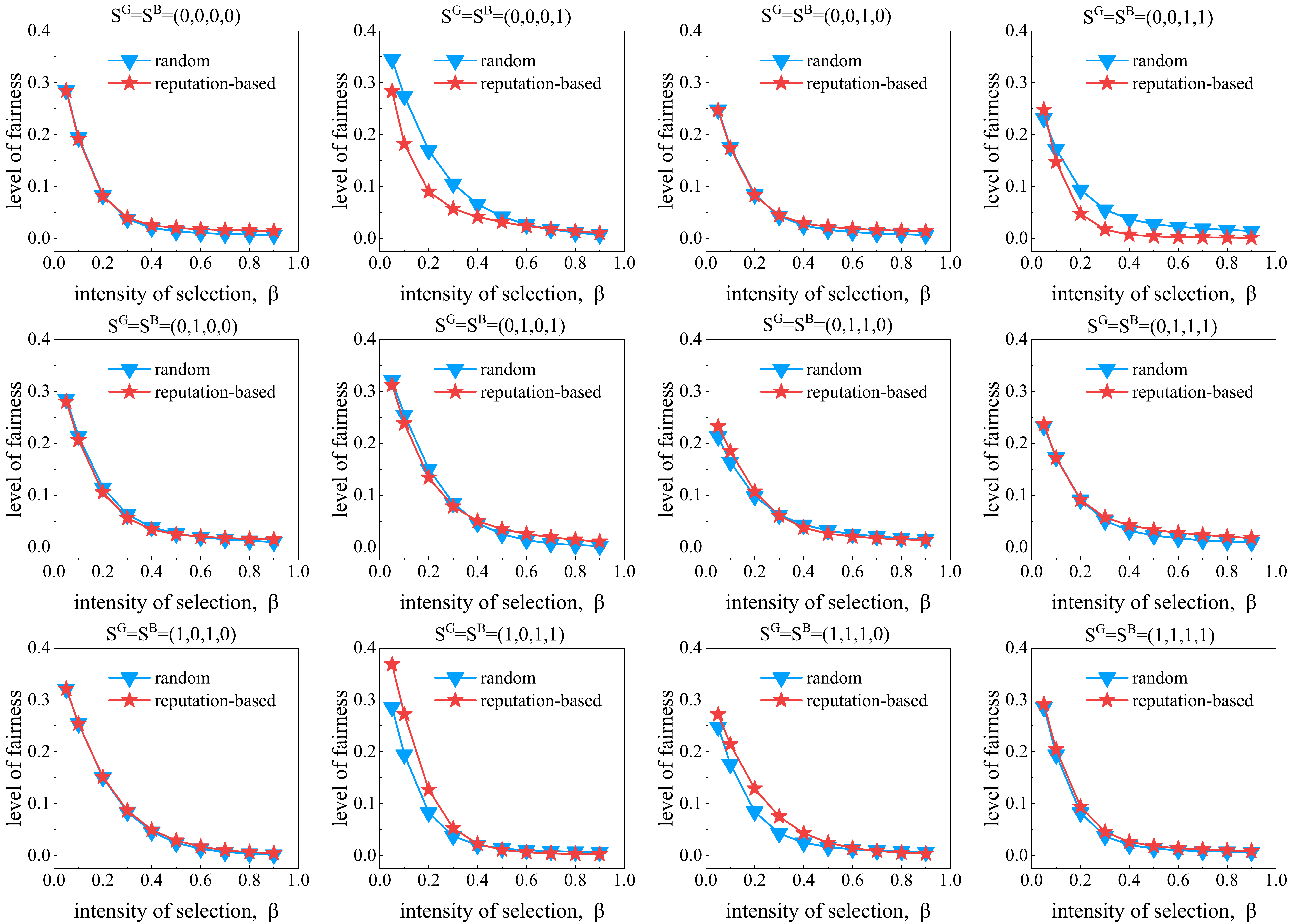}
 	\caption{The level of fairness as a function of the intensity of selection under the remaining second-order social norms. Parameters: $Z=50$, $\varepsilon=0.01$, $c_R=0.01$, and $\mu=0.01$.}
 	\label{figA1}
 \end{figure*}

\begin{figure}
	\centering
	\includegraphics[width=\linewidth]{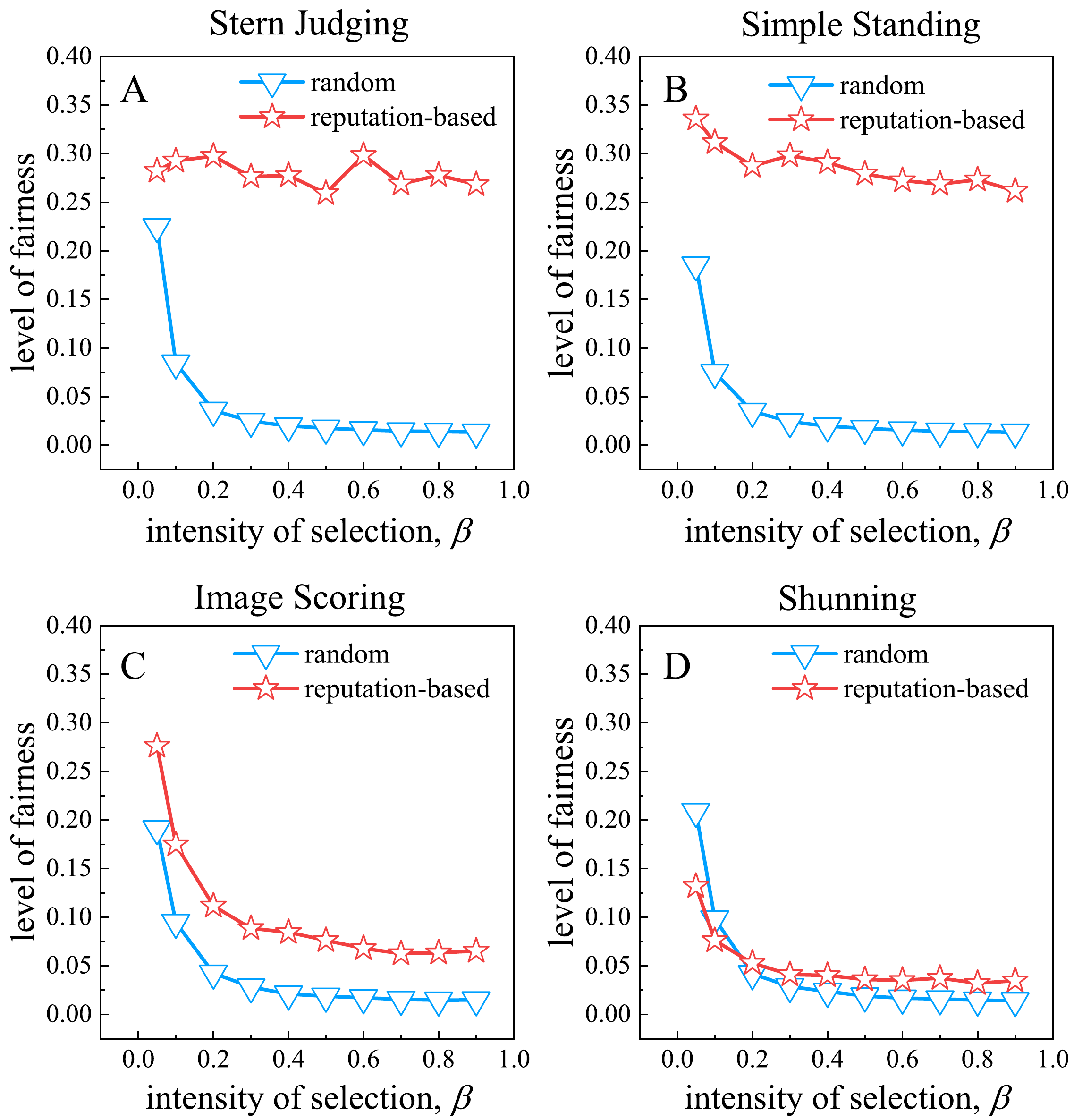}
	\caption{\label{simu} Agent-based simulations under stern judging (A), simple standing (B), image scoring (C), and shunning (D). At the beginning of each generation, each player is randomly assessed as good or bad. A generation is comprised of $5Z$ rounds. On average, a player will chosen $5$ times, either as donor or as recipient. Each point is obtained by averaging the fraction of acting fairly over $10^7$ generation. Parameters: $Z=50$, $\varepsilon=0.01$, $c_R=0.01$, $\mu=0.01$.}
\end{figure}

 \section{Reputation dynamics}
 \label{reputation}
 Assume that a population consists of $m$ players with strategy $X=s_G^Xs_B^Xs_R^X$ and $Z-m$ players using strategy $Y=s_G^Ys_B^Ys_R^Y$, and then the state of the reputation system is denoted by a two-dimensional vector $(i,j)$, which means that there are $i\in {0,1,\cdots, m}$ ($j\in {0,1,\cdots, Z-m}$) good players among $m$ $X$-players (among $Z-m$ $Y$-players). Given the reputation state is $(i,j)$ at time $t$, we can calculate the transition probability $P(i,j;i',j')$, which means how likely the population will be in state $(i',j')$ at time $t+1$. It is worth mentioning that in the pairwise competition among four silent strategies, the reputation system remains unchanged after each round of game as individuals do not report the outcome. In the remaining pairwise competition among eight strategies, $P(i,j;i',j')=0$ for $|i'-i|+|j'-j|>1$, and $P(i,j;i',j')$ is non-zero for $|i'-i|+|j'-j|\leq 1$.

 Assume the event that a $T\in \{X,Y\}$-player with reputation $M\in \{G,B\}$ reports the outcome which he just witnessed as an observer, is denoted by $Ob=TM$.
 Given the state of the reputation state is $(i,j)$, $Ob=XG$, $Ob=YG$, $Ob=XB$, or $Ob=YB$ occurs with probability
 \begin{eqnarray}
 \begin{array}{l}
 p\left( Ob=XG \right) =\frac{i}{Z}E\left( s_{R}^{X} \right), p\left( Ob=XB \right) =\frac{m-i}{Z}E\left( s_{R}^{X} \right) ,\\\\
 p\left( Ob=YG \right) =\frac{j}{Z}E\left( s_{R}^{Y} \right), p\left( Ob=YB \right) =\frac{Z-m-j}{Z}E\left( s_{R}^{Y} \right),
 \end{array}
 \end{eqnarray}
 where $E(x)=1$ if $x=R$ and $E(x)=0$ if $x=S$.
 Given that one of the four events $Ob = XG$, $XB$, $YG$, or $YB$ occurs, the conditional transition probability $(i,j)\to (i',j')$ is denoted by $p(i,j;i',j'|Ob=TM)$.
 According to the total probability theorem, we have
 \begin{eqnarray} 
 \begin{array}{l}
 p(i,j;i',j')= \sum_{TM=XB,YB,XG,YG}\\
 {p(i,j;i',j'|Ob=TM)}\times p(Ob=TM).
 \end{array}
 \end{eqnarray}

 We derive the expressions of $p(i,j;i',j'|Ob=TM)$ as follows.
 \begin{itemize}
 	\item [1.]
 	Transition $(i,j)\to (i+1,j)$ if $Ob=XB,YB,XG$ or $YG$ occurs.
 \end{itemize}

 The case occurs when a bad $X$-player is chosen as the dictator and changes his reputation from bad to good by making a fair or unfair division with a good or bad recipient.\\
 1) Two bad $X$-players are chosen with probability
 \begin{eqnarray}
 \begin{array}{ll}
 \frac{(m-i-1)(m-i-2)}{(Z-1)(Z-2)}, & \mbox{given Ob=XB};\\
 \frac{(m-i)(m-i-1)}{(Z-1)(Z-2)}, &\mbox{given Ob=YB, XG, YG}.
 \end{array}\label{iplus1}
 \end{eqnarray}
 2) A bad $X$-player and a bad $Y$-player are chosen with probability
 \begin{eqnarray}
 \begin{array}{ll}
 \frac{2(m-i-1)(Z-m-j)}{(Z-1)(Z-2)}, & \mbox{given Ob=XB};\\
 \frac{2(m-i)(Z-m-j-1)}{(Z-1)(Z-2)}, &\mbox{given Ob=YB};\\
 \frac{2(m-i)(Z-m-j)}{(Z-1)(Z-2)}, &\mbox{given Ob=XG, YG}.
 \end{array}\label{iplus2}
 \end{eqnarray}
 3) A bad $X$-player and an another good player are chosen with probability
 \begin{eqnarray}
 \begin{array}{ll}
 \frac{2(m-i-1)(i+j)}{(Z-1)(Z-2)}, & \mbox{given Ob=XB};\\
 \frac{2(m-i)(i+j)}{(Z-1)(Z-2)}, &\mbox{given Ob=YB};\\
 \frac{2(m-i)(i+j-1)}{(Z-1)(Z-2)}, &\mbox{given Ob=XG, YG}.
 \end{array}\label{iplus3}
 \end{eqnarray}
 Following the social norm $S^B=(F_G, F_B, N_G, N_B)$, the dictator switches his reputation from bad to good after a round by making a fair division or an unfair division with a good recipient with probability $I(s_G^X)(1-\varepsilon)F_G$ or $(I(s_G^X)\varepsilon +1-I(s_G^X))N_G$ ($I(x)=1$ for $x=F$ and $I(x)=0$ for $x=N$); it also occurs by making a fair or an unfair division against a bad recipient with probability $I(s_B^X)(1-\varepsilon)F_B$ or $(I(s_B^X)\varepsilon +1-I(s_B^X))N_B$.

 Combining Eqs.~(\ref{iplus1})-(\ref{iplus3}) with the random role assignment, we have the corresponding transition probabilities as
 \begin{small}
 	\begin{eqnarray}
 	\hspace{-8mm}
 	\begin{array}{l}	
 	p\left( i,j;i+1,j|Ob=XB \right)=\frac{m-i-1}{Z-1}\\
 	\left( \frac{i+j}{Z-2}\left( I\left( s_{G}^{X} \right) \left( 1-\varepsilon \right) F_G+\left( I\left( s_{G}^{X} \right) \varepsilon +1-I\left( s_{G}^{X} \right) \right) N_G \right) \right.\\
 	+\left. \frac{Z-2-i-j}{Z-2}\left( I\left( s_{B}^{X} \right) \left( 1-\varepsilon \right) F_B+\left( I\left( s_{B}^{X} \right) \varepsilon +1-I\left( s_{B}^{X} \right) \right) N_B \right) \right) ,\\
 	p\left( i,j;i+1,j|Ob=YB \right) =\frac{m-i}{Z-1}\\
 	\left( \frac{i+j}{Z-2}\left( I\left( s_{G}^{X} \right) \left( 1-\varepsilon \right) F_G+\left( I\left( s_{G}^{X} \right) \varepsilon +1-I\left( s_{G}^{X} \right) \right) N_G \right) \right.\\
 	+\left. \frac{Z-2-i-j}{Z-2}\left( I\left( s_{B}^{X} \right) \left( 1-\varepsilon \right) F_B+\left( I\left( s_{B}^{X} \right) \varepsilon +1-I\left( s_{B}^{X} \right) \right) N_B \right) \right) ,\\
 	p\left( i,j;i+1,j|Ob=XG \right) =p\left( i,j;i+1,j|Ob=YG \right) =\frac{m-i}{Z-1}\\
 	\left( \frac{i+j-1}{Z-2}\left( I\left( s_{G}^{X} \right) \left( 1-\varepsilon \right) F_G+\left( I\left( s_{G}^{X} \right) \varepsilon +1-I\left( s_{G}^{X} \right) \right) N_G \right) \right.\\
 	+\left. \frac{Z-1-i-j}{Z-2}\left( I\left( s_{B}^{X} \right) \left( 1-\varepsilon \right) F_B+\left( I\left( s_{B}^{X} \right) \varepsilon +1-I\left( s_{B}^{X} \right) \right) N_B \right) \right) .
 	\end{array}	
 	\end{eqnarray}
 \end{small}

 Combining Eqs.~(\ref{iplus1}) and (\ref{iplus2}) with the reputation-based role assignment, we have the corresponding transition probabilities as
 \begin{small}
 	\begin{eqnarray}
 	\hspace{-8mm}
 	\begin{array}{l}
 	p\left( i,j;i+1,j|Ob=XB \right) =\frac{m-i-1}{Z-1}\\
 	\frac{Z-i-j-2}{Z-2} \left( I\left( s_{B}^{X} \right) \left( 1-\varepsilon \right) F_B+\left( I\left( s_{B}^{X} \right) \varepsilon +1-I\left( s_{B}^{X} \right) \right) N_B \right) ,\\
 	p\left( i,j;i+1,j|Ob=YB \right) =\frac{m-i}{Z-1}  \\
 	\frac{Z-i-j-2}{Z-2}  \left( I\left( s_{B}^{X} \right) \left( 1-\varepsilon \right) F_B+\left( I\left( s_{B}^{X} \right) \varepsilon +1-I\left( s_{B}^{X} \right) \right) N_B \right) ,\\
 	p\left( i,j;i+1,j|Ob=XG \right) =p\left( i,j;i+1,j|Ob=YG \right) =\frac{m-i}{Z-1} \\
 	\frac{Z-i-j-1}{Z-2}
 	\left( I\left( s_{B}^{X} \right) \left( 1-\varepsilon \right) F_B+\left( I\left( s_{B}^{X} \right) \varepsilon +1-I\left( s_{B}^{X} \right) \right) N_B \right).
 	\end{array}
 	\end{eqnarray}
 \end{small}

 \begin{itemize}
 	\item [2.]
 	Transition $(i,j)\to (i-1,j)$ if $Ob=XB,YB,XG$ or $YG$ occurs.
 \end{itemize}

 The case requires that a good $X$-player is chosen as the dictator and changes his reputation from good to bad by making a fair or unfair division with a good or bad recipient.\\
 1) Two good $X$-players are chosen with probability
 \begin{eqnarray}
 \begin{array}{ll}
 \frac{(i-1)(i-2)}{(Z-1)(Z-2)},&\mbox{given Ob=XG};\\
 \frac{i(i-1)}{(Z-1)(Z-2)},&\mbox{given Ob=YG, XB, YB}.
 \end{array}\label{iminus1}
 \end{eqnarray}
 2) A good $X$-player and a good $Y$-player are chosen with probability
 \begin{eqnarray}
 \begin{array}{ll}
 \frac{2(i-1)j}{(Z-1)(Z-2)},&\mbox{given Ob=XG};\\
 \frac{2i(j-1)}{(Z-1)(Z-2)},&\mbox{given Ob=YG};\\
 \frac{2ij}{(Z-1)(Z-2)},&\mbox{given Ob=XB, YB}.
 \end{array}\label{iminus2}
 \end{eqnarray}
 3) A good $X$-player and an another bad player are chosen with probability
 \begin{eqnarray}
 \begin{array}{ll}
 \frac{2(i-1)(Z-i-j)}{(Z-1)(Z-2)},&\mbox{given Ob=XG};\\
 \frac{2i(Z-i-j)}{(Z-1)(Z-2)},&\mbox{given Ob=YG};\\
 \frac{2i(Z-i-j-1)}{(Z-1)(Z-2)},&\mbox{given Ob=XB, YB}.
 \end{array}\label{iminus3}
 \end{eqnarray}
 Following the social norm $S^G=(F_G, F_B, N_G, N_B)$, the dictator switches his reputation from good to bad after a round by making a fair division or an unfair division against a good recipient with probability $I(s_G^X)(1-\varepsilon)(1-F_G)$ or $(I(s_G^X)\varepsilon +1-I(s_G^X))(1-N_G)$; it also occurs by making a fair or an unfair division against a bad recipient with probability $I(s_B^X)(1-\varepsilon)(1-F_B)$ or $(I(s_B^X)\varepsilon +1-I(s_B^X))(1-N_B)$.

 Combining Eqs.~(\ref{iminus1})-(\ref{iminus3}) with the random role assignment, we have the corresponding conditional transition probabilities as
 \begin{small}
 	\begin{eqnarray}
 	\hspace{-8mm}
 	\begin{array}{l}
 	p\left( i,j;i-1,j|Ob=XB \right) =p\left( i,j;i-1,j|Ob=YB \right) = \frac{i}{Z-1}\\
 	\left( \frac{i+j-1}{Z-2}\left( I\left( s_{G}^{X} \right) \left( 1-\varepsilon \right) F_G^{-}+\left( I\left( s_{G}^{X} \right) \varepsilon +1-I\left( s_{G}^{X} \right) \right) N_G^{-} \right) \right.+\\\left. \frac{Z-1-i-j}{Z-2}\left( I\left( s_{B}^{X} \right) \left( 1-\varepsilon \right) F_B^{-}+\left( I\left( s_{B}^{X} \right) \varepsilon +1-I\left( s_{B}^{X} \right) \right) N_B^{-} \right) \right),\\
 	p\left( i,j;i-1,j|Ob=XG \right) =\frac{i-1}{Z-1}\\
 	\left( \frac{i+j-2}{Z-2}\left( I\left( s_{G}^{X} \right) \left( 1-\varepsilon \right) F_G^{-}+\left( I\left( s_{G}^{X} \right) \varepsilon +1-I\left( s_{G}^{X} \right) \right) N_G^{-} \right) \right.+\\
 	\left. \frac{Z-i-j}{Z-2}\left( I\left( s_{B}^{X} \right) \left( 1-\varepsilon \right) F_B^{-}+\left( I\left( s_{B}^{X} \right) \varepsilon +1-I\left( s_{B}^{X} \right) \right) N_B^{-} \right) \right),\\
 	p\left( i,j;i-1,j|Ob=YG \right) = \frac{i}{Z-1}\\
 	\left( \frac{i+j-2}{Z-2}\left( I\left( s_{G}^{X} \right) \left( 1-\varepsilon \right) F_G^{-}+\left( I\left( s_{G}^{X} \right) \varepsilon +1-I\left( s_{G}^{X} \right) \right) N_G^{-} \right) \right.+\\
 	\left. \frac{Z-i-j}{Z-2}\left( I\left( s_{B}^{X} \right) \left( 1-\varepsilon \right) F_B^{-}+\left( I\left( s_{B}^{X} \right) \varepsilon +1-I\left( s_{B}^{X} \right) \right) N_B^{-} \right) \right) .
 	\end{array}
 	\end{eqnarray}
 \end{small}
 where $F_G^{-}=1-F_G$, $F_B^{-}=1-F_B$, $N_G^{-}=1-N_G$, and $N_B^{-}=1-N_B$.

 Combining Eqs.~(\ref{iminus1})-(\ref{iminus3}) with the reputation-based role assignment, we have the corresponding transition probabilities as
 \begin{small}
 	\begin{eqnarray}
 	\hspace{-8mm}
 	\begin{array}{l}
 	p\left( i,j;i-1,j|Ob=XB \right) =p\left( i,j;i-1,j|Ob=YB \right) =\frac{i}{Z-1}\\
 	\left( \frac{i+j-1}{Z-2}\left( I\left( s_{G}^{X} \right) \left( 1-\varepsilon \right) F_G^{-}+\left( I\left( s_{G}^{X} \right) \varepsilon +1-I\left( s_{G}^{X} \right) \right) N_G^{-} \right) \right. +\\
 	\left. \frac{2\left( Z-1-i-j \right)}{Z-2}\left( I\left( s_{B}^{X} \right) \left( 1-\varepsilon \right) F_B^{-}+\left( I\left( s_{B}^{X} \right) \varepsilon +1-I\left( s_{B}^{X} \right) \right) N_B^{-} \right) \right),\\
 	p\left( i,j;i-1,j|Ob=XG \right) =\frac{i-1}{Z-1}\\
 	\left( \frac{i+j-2}{Z-2}\left( I\left( s_{G}^{X} \right) \left( 1-\varepsilon \right) F_G^{-}+\left( I\left( s_{G}^{X} \right) \varepsilon +1-I\left( s_{G}^{X} \right) \right) N_G^{-} \right) \right.+\\
 	\left. \frac{2\left( Z-i-j \right)}{Z-2}\left( I\left( s_{B}^{X} \right) \left( 1-\varepsilon \right) F_B^{-}+\left( I\left( s_{B}^{X} \right) \varepsilon +1-I\left( s_{B}^{X} \right) \right) N_B^{-} \right) \right),\\
 	p\left( i,j;i-1,j|Ob=YG \right) =\frac{i}{Z-1}\\
 	\left( \frac{i+j-2}{Z-2}\left( I\left( s_{G}^{X} \right) \left( 1-\varepsilon \right) F_G^{-}+\left( I\left( s_{G}^{X} \right) \varepsilon +1-I\left( s_{G}^{X} \right) \right) N_G^{-} \right) \right.+\\
 	\left. \frac{2\left(Z-i-j \right)}{Z-2}\left( I\left( s_{B}^{X} \right) \left( 1-\varepsilon \right) F_B^{-}+\left( I\left( s_{B}^{X} \right) \varepsilon +1-I\left( s_{B}^{X} \right) \right) N_B^{-} \right) \right).
 	\end{array}	
 	\end{eqnarray}
 \end{small}

 \begin{itemize}
 	\item [3.]
 	Transition $(i,j)\to (i,j+1)$ if $Ob=XB,YB,XG$ or $YG$ occurs.
 \end{itemize}
 The case occurs when a bad $Y$-player is chosen as the dictator and changes his reputation from bad to good by making a fair or unfair division with a good or bad recipient.\\
 1) Two bad $Y$-players are chosen with probability\\
 \begin{eqnarray}
 \begin{array}{ll}
 \frac{(Z-m-j-1)(Z-m-j-2)}{(Z-1)(Z-2)},&\mbox{given Ob=YB};\\
 \frac{(Z-m-j)(Z-m-j-1)}{(Z-1)(Z-2)}, &\mbox{given Ob=XB, XG, YG}.\\
 \end{array}\label{jplus1}
 \end{eqnarray}
 2) A bad $Y$-player and a bad $X$-player are chosen with probability
 \begin{eqnarray}
 \begin{array}{ll}
 \frac{2(Z-m-j-1)(m-i)}{(Z-1)(Z-2)},&\mbox{given Ob=YB};\\
 \frac{2(Z-m-j)(m-i-1)}{(Z-1)(Z-2)}, &\mbox{given Ob=XB};\\
 \frac{2(Z-m-j)(m-i)}{(Z-1)(Z-2)},&\mbox{given Ob=XG, YG}.
 \end{array}\label{jplus2}
 \end{eqnarray}
 3) A bad $Y$-player and an another good player are chosen with probability
 \begin{eqnarray}
 \begin{array}{ll}
 \frac{2(Z-m-j-1)(i+j)}{(Z-1)(Z-2)},&\mbox{given Ob=YB};\\
 \frac{2(Z-m-j)(i+j)}{(Z-1)(Z-2)}, &\mbox{given Ob=XB};\\
 \frac{2(Z-m-j)(i+j-1)}{(Z-1)(Z-2)},&\mbox{given Ob=XG, YG}.
 \end{array}\label{jplus3}
 \end{eqnarray}
 Following the social norm $S^B=(F_G, F_B, N_G, N_B)$, the dictator switches his reputation from bad to good after a round by making a fair division or an unfair division against a good recipient with probability $I(s_G^Y)(1-\varepsilon)F_G$ or $(I(s_G^Y)\varepsilon +1-I(s_G^Y))N_G$; it also occurs by making a fair or an unfair division against a bad recipient with probability $I(s_B^Y)(1-\varepsilon)F_B$ or $(I(s_B^Y)\varepsilon +1-I(s_B^Y))N_B$.

 Combining Eqs.~(\ref{jplus1})-(\ref{jplus3}) with the random role assignment, we have the corresponding conditional transition probabilities as
 \begin{small}
 	\begin{eqnarray}  
 	\hspace{-8mm}
 	\begin{array}{l}		
 	p\left( i,j;i,j+1|Ob=XB \right) = \frac{Z-m-j}{Z-1}\\
 	\left( \frac{i+j}{Z-2}\left( I\left( s_{G}^{Y} \right) \left( 1-\varepsilon \right) F_G+\left( I\left( s_{G}^{Y} \right) \varepsilon +1-I\left( s_{G}^{Y} \right) \right) N_G \right) \right.+\\
 	\left. \frac{Z-2-i-j}{Z-2}\left( I\left( s_{B}^{Y} \right) \left( 1-\varepsilon \right) F_B+\left( I\left( s_{B}^{Y} \right) \varepsilon +1-I\left( s_{B}^{Y} \right) \right) N_B \right) \right),\\
 	p\left( i,j;i,j+1|Ob=YB \right) =\frac{Z-m-j-1}{Z-1}\\
 	\left( \frac{i+j}{Z-2}\left( I\left( s_{G}^{Y} \right) \left( 1-\varepsilon \right) F_G+\left( I\left( s_{G}^{Y} \right) \varepsilon +1-I\left( s_{G}^{Y} \right) \right) N_G \right) \right.+\\
 	\left. \frac{Z-2-i-j}{Z-2}\left( I\left( s_{B}^{Y} \right) \left( 1-\varepsilon \right) F_B+\left( I\left( s_{B}^{Y} \right) \varepsilon +1-I\left( s_{B}^{Y} \right) \right) N_B \right) \right),\\
 	p\left( i,j;i,j+1|Ob=XG \right) \!\!=\!\!p\left( i,j;i,j+1|Ob=YG \right)\!\!=\!\!\frac{Z-m-j}{Z-1}\\
 	\left( \frac{i+j-1}{Z-2}\left( I\left( s_{G}^{Y} \right) \left( 1-\varepsilon \right) F_G+\left( I\left( s_{G}^{Y} \right) \varepsilon +1-I\left( s_{G}^{Y} \right) \right) N_G \right) \right.+\\
 	\left. \frac{Z-1-i-j}{Z-2}\left( I\left( s_{B}^{Y} \right) \left( 1-\varepsilon \right) F_B+\left( I\left( s_{B}^{Y} \right) \varepsilon +1-I\left( s_{B}^{Y} \right) \right) N_B \right) \right).
 	\end{array}
 	\end{eqnarray}
 \end{small}

 Combining Eqs.~(\ref{jplus1}) and (\ref{jplus2}) with the reputation-based role assignment, we have the corresponding transition probabilities as
 \begin{small}
 	\begin{eqnarray} 
 	\hspace{-8mm}
 	\begin{array}{l}
 	p\left( i,j;i,j+1|Ob=XB \right) =\frac{Z-m-j}{Z-1}\\
 	\frac{Z-i-j-2}{Z-2} \left( I\left( s_{B}^{Y} \right) \left( 1-\varepsilon \right) F_B+\left( I\left( s_{B}^{Y} \right) \varepsilon +1-I\left( s_{B}^{Y} \right) \right) N_B \right),\\
 	p\left( i,j;i,j+1|Ob=YB \right) =\frac{Z-m-j-1}{Z-1}\\
 	\frac{Z-i-j-2}{Z-2} \left( I\left( s_{B}^{Y} \right) \left( 1-\varepsilon \right) F_B+\left( I\left( s_{B}^{Y} \right) \varepsilon +1-I\left( s_{B}^{Y} \right) \right) N_B \right),\\
 	p\left( i,j;i,j+1|Ob=XG \right)\!\!=\!\!p\left( i,j;i+1,j|Ob=YG \right)\!\! =\!\!\frac{Z-m-j}{Z-1}\\
 	\frac{Z-i-j-1}{Z-2} \left( I\left( s_{B}^{Y} \right) \left( 1-\varepsilon \right) F_B+\left( I\left( s_{B}^{Y} \right) \varepsilon +1-I\left( s_{B}^{Y} \right) \right) N_B \right).
 	\end{array}
 	\end{eqnarray}
 \end{small}

 \begin{itemize}
 	\item [4.]
 	Transition $(i,j)\to (i,j-1)$ if $Ob=XB,YB,XG$ or $YG$ occurs.
 \end{itemize}
 The case occurs when a good $Y$-player is chosen as the dictator and changes his reputation from good to bad by making a fair or unfair division with a good or bad recipient.\\
 1) Two good $Y$-players are chosen with probability
 \begin{eqnarray}
 \begin{array}{ll}
 \frac{(j-1)(j-2)}{(Z-1)(Z-2)},&\mbox{given Ob=YG};\\
 \frac{j(j-1)}{(Z-1)(Z-2)}, &\mbox{given Ob=XG, XB, YB}.\\
 \end{array}\label{jminus1}
 \end{eqnarray}
 2) A good $Y$-player and a good $X$-player are chosen with probability
 \begin{eqnarray}
 \begin{array}{ll}
 \frac{2(j-1)i}{(Z-1)(Z-2)},&\mbox{given Ob=YG};\\
 \frac{2j(i-1)}{(Z-1)(Z-2)}, &\mbox{given Ob=XG};\\
 \frac{2ij}{(Z-1)(Z-2)}, & \mbox{given Ob=XB, YB}.
 \end{array}\label{jminus2}
 \end{eqnarray}
 3) A good $Y$-player and an another bad player are chosen with probability
 \begin{eqnarray}
 \begin{array}{ll}
 \frac{2(j-1)(Z-i-j)}{(Z-1)(Z-2)},&\mbox{given Ob=YG};\\
 \frac{2j(Z-i-j)}{(Z-1)(Z-2)}, &\mbox{given Ob=XG};\\
 \frac{2j(Z-i-j-1)}{(Z-1)(Z-2)}, & \mbox{given Ob=XB, YB}.
 \end{array}\label{jminus3}
 \end{eqnarray}
 Following the social norm $S^G=(F_G, F_B, N_G, N_B)$, the dictator switches his reputation from good to bad after a round by making a fair division or an unfair division against a good recipient with probability $I(s_G^Y)(1-\varepsilon)(1-F_G)$ or $(I(s_G^Y)\varepsilon +1-I(s_G^Y))(1-N_G)$; it also occurs by making a fair or an unfair division against a bad recipient with probability $I(s_B^Y)(1-\varepsilon)(1-F_B)$ or $(I(s_B^Y)\varepsilon +1-I(s_B^Y))(1-N_B)$.

 Combining Eqs.~(\ref{jminus1})-(\ref{jminus3}) with the random role assignment, we have the corresponding conditional transition probabilities as
 \begin{small}
 	\begin{eqnarray} 
 	\hspace{-8mm}
 	\begin{array}{l}	
 	p\left( i,j;i,j-1|Ob=XB \right) =\left( i,j;i,j-1|Ob=YB \right) = \frac{j}{Z-1}\\
 	\left( \frac{i+j-1}{Z-2}\left( I\left( s_{G}^{Y} \right) \left( 1-\varepsilon \right) F_G^{-}+\left( I\left( s_{G}^{Y} \right) \varepsilon +1-I\left( s_{G}^{Y} \right) \right) N_G^{-} \right) \right.\\
 	+\left. \frac{Z-1-i-j}{Z-2}\left( I\left( s_{B}^{Y} \right) \left( 1-\varepsilon \right) F_B^{-}+\left( I\left( s_{B}^{Y} \right) \varepsilon +1-I\left( s_{B}^{Y} \right) \right) N_B^{-} \right) \right),\\
 	p\left( i,j;i,j-1|Ob=XG \right) =\frac{j}{Z-1}\\
 	\left( \frac{i+j-2}{Z-2}\left( I\left( s_{G}^{Y} \right) \left( 1-\varepsilon \right) F_G^{-}+\left( I\left( s_{G}^{Y} \right) \varepsilon +1-I\left( s_{G}^{Y} \right) \right) N_G^{-} \right) \right.\\
 	+\left. \frac{Z-i-j}{Z-2}\left( I\left( s_{B}^{Y} \right) \left( 1-\varepsilon \right) F_B^{-}+\left( I\left( s_{B}^{Y} \right) \varepsilon +1-I\left( s_{B}^{Y} \right) \right) N_B^{-} \right) \right),\\
 	p\left( i,j;i,j-1|Ob=YG \right) =\frac{j-1}{Z-1}\\
 	\left( \frac{i+j-2}{Z-2}\left( I\left( s_{G}^{Y} \right) \left( 1-\varepsilon \right) F_G^{-}+\left( I\left( s_{G}^{Y} \right) \varepsilon +1-I\left( s_{G}^{Y} \right) \right) N_G^{-} \right) \right.\\
 	+\left. \frac{Z-i-j}{Z-2}\left( I\left( s_{B}^{Y} \right) \left( 1-\varepsilon \right) F_B^{-}+\left( I\left( s_{B}^{Y} \right) \varepsilon +1-I\left( s_{B}^{Y} \right) \right) N_B^{-} \right) \right).
 	\end{array}
 	\end{eqnarray}
 \end{small}
 where $F_G^{-}=1-F_G$, $F_B^{-}=1-F_B$, $N_G^{-}=1-N_G$, and $N_B^{-}=1-N_B$.

 Combining Eqs.~(\ref{jminus1}) and (\ref{jminus3}) with the reputation-based role assignment, we have the corresponding transition probabilities as
 \begin{small}
 	\begin{eqnarray} 
 	\hspace{-8mm}
 	\begin{array}{l}
 	p\left( i,j;i,j-1|Ob=XB \right) =p\left( i,j;i,j-1|Ob=YB \right) =\frac{j}{Z-1}\\
 	\left(  \frac{i+j-1}{Z-2} \left( I\left( s_{G}^{Y} \right) \left( 1-\varepsilon \right) F_G^{-}+\left( I\left( s_{G}^{Y} \right) \varepsilon +1-I\left( s_{G}^{Y} \right) \right) N_G^{-} \right) \right.\\
 	+\left. \frac{2(Z-1-i-j)}{Z-2}\left( I\left( s_{B}^{Y} \right) \left( 1-\varepsilon \right) F_B^{-}+\left( I\left( s_{B}^{Y} \right) \varepsilon +1-I\left( s_{B}^{Y} \right) \right) N_B^{-} \right) \right),\\
 	p\left( i,j;i,j-1|Ob=XG \right) =\frac{j}{Z-1}\\
 	\left(   \frac{i+j-2}{Z-2} \left( I\left( s_{G}^{Y} \right) \left( 1-\varepsilon \right) F_G^{-}+\left( I\left( s_{G}^{Y} \right) \varepsilon +1-I\left( s_{G}^{Y} \right) \right) N_G^{-} \right) \right.\\
 	+\left. \frac{2(Z-i-j)}{Z-2}\left( I\left( s_{B}^{Y} \right) \left( 1-\varepsilon \right) F_B^{-}+\left( I\left( s_{B}^{Y} \right) \varepsilon +1-I\left( s_{B}^{Y} \right) \right) N_B^{-} \right) \right),\\
 	p\left( i,j;i,j-1|Ob=YG \right) =\frac{j-1}{Z-1}\\
 	\left(   \frac{i+j-2}{Z-2} \left( I\left( s_{G}^{Y} \right) \left( 1-\varepsilon \right) F_G^{-}+\left( I\left( s_{G}^{Y} \right) \varepsilon +1-I\left( s_{G}^{Y} \right) \right) N_G^{-} \right) \right.\\
 	+\left. \frac{2(Z-i-j)}{Z-2}\left( I\left( s_{B}^{Y} \right) \left( 1-\varepsilon \right) F_B^{-}+\left( I\left( s_{B}^{Y} \right) \varepsilon +1-I\left( s_{B}^{Y} \right) \right) N_B^{-} \right) \right).
 	\end{array}
 	\end{eqnarray}
 \end{small}

 \begin{itemize}
 	\item [5.]
 	Transition $(i,j)\to (i,j)$ if $Ob=XB,YB,XG$ or $YG$ occurs.
 \end{itemize}
 Here, the transition $(i,j)\to (i',j')$ occurs with a non-zero probability only for $|i'-i|+|j'-j|\leq 1$. Hence we have
 \begin{small}
 	
 	\begin{eqnarray} 
 	\hspace{-13mm}
 	\begin{array}{l}
 	p(i,j;i,j|Ob=TM)=\\
 	1-p(i,j;i+1,j|Ob=TM)-p(i,j;i-1,j|Ob=TM) \\
 	-p(i,j;i,j+1|Ob=TM)-p(i,j;i,j-1|Ob=TM).
 	\end{array}
 	\end{eqnarray}
 \end{small}

 \section{Expected payoffs}
 \label{payoff}
 The expression of $\pi _X( m,i,j )$ or $\pi _Y( m,i,j )$  includes four parts: the first two parts denote
 the expected payoff of an $X$ or $Y$ player in the role of dictator by making a fair split and by making an unfair split, respectively; the third part is the expected payoff of an $X$ or $Y$ player in the role of recipient by receiving a fair split; the fourth part is the cost in the role of observer.

 In the random role assignment, the two participants have the same possibility of becoming the dictator. Then we have
 \begin{small}
 	\hspace{-30mm}
 	\begin{eqnarray}
 	\begin{array}{l}
 	\pi _X\left( m,i,j \right) =\\
 	\frac{1}{2}(1-\epsilon )\left( \frac{i}{m}\left( 0.5*I\left( s_{G}^{X} \right) \frac{i+j-1}{Z-1}+0.5*I\left( s_{B}^{X} \right) \frac{Z-i-j}{Z-1} \right) \right. +\\
 	\qquad	\qquad\left. \frac{m-i}{m}\left( 0.5*I\left( s_{G}^{X} \right) \frac{i+j}{Z-1}+0.5*I\left( s_{B}^{X} \right) \frac{Z-1-i-j}{Z-1} \right) \right)+\\
 	1 \left( \frac{i}{m}0.5* \left( 1-I\left( s_{G}^{X} \right) +\epsilon I\left( s_{G}^{X} \right)
 	\right) \frac{i+j-1}{Z-1} \right.+\\ 	
 	\,\,\,\,\,\,\,\,\,\,\frac{i}{m}0.5*\left( 1-I\left( s_{B}^{X} \right) +\epsilon I\left( s_{B}^{X} \right) \right) \frac{Z-i-j}{Z-1} +\\
 	\,\,\,\,\,\,\,\,\,\, \frac{m-i}{m}0.5* \left( 1-I\left( s_{G}^{X} \right)+\epsilon I\left( s_{G}^{X} \right) \right) \frac{i+j}{Z-1}+\\ 	
 	\,\,\,\,\,\,\,\,\,\,\left.\frac{m-i}{m}0.5*\left(1-I\left( s_{B}^{X} \right) +\epsilon I\left( s_{B}^{X} \right)\right) \frac{Z-1-i-j}{Z-1}  \right)+\\
 	\frac{1}{2}(1-\epsilon )\left( \frac{i}{m}0.5*\left( I\left( s_{G}^{X} \right) \frac{m-1}{Z-1}+I\left( s_{G}^{Y} \right) \frac{Z-m}{Z-1} \right) +\right.\\
 	\qquad\qquad \left.\frac{m-i}{m}0.5*\left( I\left( s_{B}^{X} \right) \frac{m-1}{Z-1}+I\left( s_{B}^{Y} \right) \frac{Z-m}{Z-1} \right) \right) \\
 	-c_RE\left( s_{R}^{X} \right),
 	\end{array}
 	\end{eqnarray}
 \end{small}
 \begin{small}
 	\hspace{-8mm}
 	\begin{eqnarray}
 	\begin{array}{l}
 	\pi _Y\left( m,i,j \right) = \\
 	\frac{1}{2}(1-\epsilon )\left( \frac{j}{Z-m}\left(0.5* I\left( s_{G}^{Y} \right) \frac{i+j-1}{Z-1}+0.5*I\left( s_{B}^{Y} \right) \frac{Z-i-j}{Z-1} \right) + \right.
 	\\
 	\qquad\qquad	\left. \frac{Z-m-j}{Z-m}\left( 0.5*I\left( s_{G}^{Y} \right) \frac{i+j}{Z-1}+0.5*I\left( s_{B}^{Y} \right) \frac{Z-1-i-j}{Z-1} \right) \right) +
 	\\
 	1\times \left( \frac{j}{Z-m}0.5* \left( 1-I\left( s_{G}^{Y} \right) +
 	\epsilon I\left( s_{G}^{Y} \right)\right) \frac{i+j-1}{Z-1}+\right.\\
 	\qquad \frac{j}{Z-m}0.5* \left( 1-I\left( s_{B}^{Y} \right) +\epsilon I\left( s_{B}^{Y} \right)\right) \frac{Z-i-j}{Z-1}  +\\
 	\qquad \frac{Z-m-j}{Z-m}0.5*\left( 1-I\left( s_{G}^{Y} \right) +\epsilon I\left(s_{G}^{Y} \right) \right) \frac{i+j}{Z-1}+\\
 	\qquad\left. \frac{Z-m-j}{Z-m}0.5*\left(1-I\left( s_{B}^{Y} \right)+\epsilon I\left( s_{B}^{Y} \right) \right) \frac{Z-1-i-j}{Z-1}\right) +
 	\\
 	\frac{1}{2}(1-\epsilon )\left( \frac{j}{Z-m}0.5*\left( I\left( s_{G}^{X} \right) \frac{m}{Z-1}+I\left( s_{G}^{Y} \right) \frac{Z-m-1}{Z-1} \right) \right. +
 	\\
 	\qquad\qquad\left. \frac{Z-m-j}{Z-m}0.5*\left( I\left( s_{B}^{X} \right) \frac{m}{Z-1}+I\left( s_{B}^{Y} \right) \frac{Z-m-1}{Z-1} \right) \right)
 	\\
 	-c_RE\left( s_{R}^{Y} \right),
 	\end{array}
 	\end{eqnarray}
 \end{small}

 In the reputation-based role assignment, the individual with good reputation plays the role of dictator when the opponent is a bad individual; two players with the same reputation are randomly chosen as the dictator or the recipient. Then, we have

 \begin{small}
 	\hspace{-30mm}
 	\begin{eqnarray}
 	\begin{array}{l}
 	\pi _X\left( m,i,j \right) =\\
 	\frac{1}{2}(1-\epsilon )\left( \frac{i}{m}\left( 0.5*I\left( s_{G}^{X} \right) \frac{i+j-1}{Z-1}+I\left( s_{B}^{X} \right) \frac{Z-i-j}{Z-1} \right) \right. +\\
 	\qquad	\qquad\left. \frac{m-i}{m} 0.5*I\left( s_{B}^{X} \right) \frac{Z-1-i-j}{Z-1} \right) +\\
 	1 \left( \frac{i}{m}0.5* \left( 1-I\left( s_{G}^{X} \right) +\epsilon I\left( s_{G}^{X} \right)
 	\right) \frac{i+j-1}{Z-1} \right.+\\ 	
 	\,\,\,\,\,\,\,\,\,\,\frac{i}{m}\left( 1-I\left( s_{B}^{X} \right) +\epsilon I\left( s_{B}^{X} \right) \right) \frac{Z-i-j}{Z-1} +\\	
 	\,\,\,\,\,\,\,\,\,\,\left.\frac{m-i}{m}0.5*\left(1-I\left( s_{B}^{X} \right) +\epsilon I\left( s_{B}^{X} \right)\right) \frac{Z-1-i-j}{Z-1}  \right)+\\
 	\frac{1}{2}(1-\epsilon )\left( \frac{i}{m}0.5*\left( I\left( s_{G}^{X} \right) \frac{i-1}{m-1}+I\left( s_{G}^{Y} \right) \frac{j}{Z-m} \right) +\right.\\
 	\qquad\qquad \frac{m-i}{m}0.5*\left( I\left( s_{B}^{X} \right) \frac{m-i-1}{m-1}+I\left( s_{B}^{Y} \right) \frac{Z-m-j}{Z-m} \right)+ \\
 	\qquad\qquad \left.\frac{m-i}{m}\left( I\left( s_{B}^{X} \right) \frac{i}{m-1}+I\left( s_{B}^{Y} \right) \frac{j}{Z-m} \right) \right) \\
 	-c_RE\left( s_{R}^{X} \right),
 	\end{array}
 	\end{eqnarray}
 \end{small}

 \begin{small}
 	\hspace{-8mm}
 	\begin{eqnarray}
 	\begin{array}{l}
 	\pi _Y\left( m,i,j \right) = \\
 	\frac{1}{2}(1-\epsilon )\left( \frac{j}{Z-m}\left(0.5* I\left( s_{G}^{Y} \right) \frac{i+j-1}{Z-1}+I\left( s_{B}^{Y} \right) \frac{Z-i-j}{Z-1} \right) + \right.
 	\\
 	\qquad\qquad	\left. \frac{Z-m-j}{Z-m}0.5*I\left( s_{B}^{Y} \right) \frac{Z-1-i-j}{Z-1} \right) +
 	\\
 	1\times \left( \frac{j}{Z-m}0.5* \left( 1-I\left( s_{G}^{Y} \right) +
 	\epsilon I\left( s_{G}^{Y} \right)\right) \frac{i+j-1}{Z-1}+\right.\\
 	\qquad \frac{j}{Z-m}\left( 1-I\left( s_{B}^{Y} \right) +\epsilon I\left( s_{B}^{Y} \right)\right) \frac{Z-i-j}{Z-1}  +\\
 	\qquad \left.\frac{Z-m-j}{Z-m}0.5*\left(1-I\left( s_{B}^{Y} \right)+\epsilon I\left( s_{B}^{Y} \right) \right) \frac{Z-1-i-j}{Z-1}\right) +
 	\\
 	\frac{1}{2}(1-\epsilon )\left( \frac{j}{Z-m}0.5*\left( I\left( s_{G}^{X} \right) \frac{i}{m}+I\left( s_{G}^{Y} \right) \frac{j-1}{Z-m-1} \right) \right. +
 	\\
 	\qquad\qquad \frac{Z-m-j}{Z-m}0.5*\left( I\left( s_{B}^{X} \right) \frac{m-i}{m}+I\left( s_{B}^{Y} \right) \frac{Z-m-j-1}{Z-m-1} \right) +
 	\\
 	\qquad\qquad \left.\frac{Z-m-j}{Z-m}\left( I\left( s_{B}^{X} \right) \frac{i}{m}+I\left( s_{B}^{Y} \right) \frac{j}{Z-m-1} \right)\right) \\
 	-c_RE\left( s_{R}^{Y} \right),
 	\end{array}
 	\end{eqnarray}
 \end{small}

 \section{Stochastic matrix of the strategy dynamics}
 \label{fixation}

 Under the scenario of rare mutations($\mu \to 0$), the population consists of almost one or two strategies
 all the time, and a mutant either fixates in the population before a new mutant appears or dies out from the population; furthermore, the monomorphic states that all players apply identical strategy predominates in the interval between two mutant events, and the polymorphic states with two different strategies are transient.
 We can describe the strategy dynamics by an embedded Markov Chain, using all eight monomorphic states as state space and fixation probabilities among those strategies for the transition probabilities. The corresponding stochastic matrix $A$ of the strategy dynamics is sized $8\times 8$ whose each entry $a_{XY}=\frac{\mu}{8} \rho_{XY}$ if $X\ne Y, a_{XX}=1-\begin{matrix} \sum_{Y,Y\ne X} a_{YX} \end{matrix}$ otherwise.

 Let $\rho_{XY}$ denote the probability that a single $X$-mutant fixates in a population with residents equipped with strategy $Y$.
 The expression of fixation probability is thus given as
 \begin{eqnarray} 
 \rho _{XY}=\left( 1+\sum_{k=1}^{Z-1}{\prod_{m=1}^k{\frac{T^-\left( m \right)}{T^+\left( m \right)}}} \right) ^{-1},
 \end{eqnarray}
 where $T^-(m)$ or $T^+(m)$ is the probability to change the number of $X$-players by $-1$ or $+1$ when the population consists of $m$ $X$-players and $Z-m$ $Y$-players.
 The number of $X$-players increases(or decreases) when a $Y$-player(or an $X$-player) is chosen to imitate an $X$-player(or a $Y$-player), accordingly we have
 \begin{eqnarray}
 \begin{array}{l}
 T^+(m)=\frac{Z-m}{Z}\frac{m}{Z-1}\frac{1}{1+e^{-\beta (g_X(m)-g_Y(m))}}, \\
 T^-(m)=\frac{m}{Z}\frac{Z-m}{Z-1}\frac{1}{1+e^{-\beta (g_Y(m)-g_X(m))}}.
 \end{array}
 \end{eqnarray}

		\nocite{*}
		\bibliography{ref}
		
	\end{document}